\documentclass[useAMS,usenatbib]{mn2e}

\usepackage{epsfig,lscape} 
\usepackage{hyperref} 
\usepackage{graphics}
\usepackage{booktabs}
\usepackage[caption=false]{subfig}
\usepackage{textcomp}
\usepackage{xcolor}
\usepackage{arydshln}
\usepackage{natbib}
\usepackage{amsmath}
\usepackage{comment}

\newcommand{\mnras}{MNRAS}
\newcommand{\apj}{ApJ}

\newcommand{\aap}{A\&A}
\newcommand{\aj}{AJ}
\newcommand{\apjl}{ApJL}
\newcommand{\apjs}{ApJS}
\newcommand{\pasp}{PASP}
\newcommand{\ao}{ApOpt}
\title[Milky Way's bulge SFH]{The Milky Way's bulge star formation history as constrained from its bimodal chemical abundance distribution}

\author[J. Lian et al.]{{Jianhui~Lian}$^{1}$\thanks{jianhui.lian@astro.utah.edu},
{Gail Zasowski}$^{1}$, {Sten Hasselquist}$^{1}$,
{David M. Nataf}$^{2}$, {Daniel Thomas}$^{3}$, \newauthor {Christian Moni Bidin}$^{4}$,
{Jos\'e G. Fern\'andez-Trincado}$^{5}$, {D. A. Garcia-Hernandez}$^{6,7}$, 
\newauthor {Richard R. Lane}$^{8}$, {Steven R. Majewski}$^{9}$, {Alexandre Roman-Lopes}$^{10}$, 
\newauthor {Mathias Schultheis}$^{11}$ \\
\small $^{1}${Department of Physics \& Astronomy, University of Utah, Salt Lake City, UT 84112, USA} \\
\small $^{2}${Center for Astrophysical Sciences and Department of Physics and Astronomy, The Johns Hopkins University, Baltimore, MD 21218} \\
\small $^{3}${Institute of Cosmology and Gravitation, University of Portsmouth, Burnaby Road, Portsmouth, UK, PO1 3FX} \\
\small $^{4}${Universidad Cat\'olica del Norte, Instituto de Astronom\'ia, Av. Angamos 0610, Antofagasta, Chile}\\
\small $^{5}${Instituto de Astronom\'ia y Ciencias Planetarias, Universidad de Atacama, Copayapu 485, Copiap\'o, Chile}\\
\small $^{6}${Instituto de Astrofısica de Canarias, 38205 La Laguna, Tenerife, Spain}\\
\small $^{7}${Universidad de La Laguna (ULL), Departamento de Astrofísica, E-38206 La Laguna, Tenerife, Spain}\\
\small $^{8}${Instituto de Astrofısica, Pontificia Universidad Cat\'olica de Chile, Av. Vicuna Mackenna 4860, 782-0436 Macul, Santiago, Chile}\\
\small $^{9}${Department of Astronomy, University of Virginia, Charlottesville, VA 22904-4325, USA}\\
\small $^{10}${Departamento de Física, Facultad de Ciencias, Universidad de La Serena, Cisternas 1200, La Serena, Chile}\\
\small $^{11}${Observatoire de la C\^ote d'Azur, Laboratoire Lagrange, 06304 Nice Cedex 4, France}\\
}
\begin{document}
\maketitle

\begin{abstract}	
%The star formation history of Galactic stellar components are well constrained by the observed chemical abundance distribution. 
We conduct a quantitative analysis of the star formation history (SFH) of the Milky Way's bulge by exploiting the constraining power of its stellar [Fe/H] and [Mg/Fe] distribution functions. 
Using APOGEE data, we confirm the previously-established bimodal [Mg/Fe]--[Fe/H] distribution within 3$\;$kpc of the inner Galaxy. %The high-$\alpha$ population peaks at a slightly higher [Mg/Fe] compared to that in the solar vicinity ($\triangle {\rm [Mg/Fe]}\sim0.05\;$dex) and is more extended towards intermediate [Mg/Fe] with negative skewness.  
Compared to that in the solar vicinity, the high-$\alpha$ population in the bulge peaks at a lower [Fe/H]. 
%has a wider distribution in both [Mg/Fe] and [Fe/H].  %with negative skewness.  
To fit these observations, we use a simple but flexible star formation framework, which assumes two distinct stages of gas accretion and star formation, and systematically evaluate a wide multi-dimensional parameter space. 
%develop a simple but flexible star formation framework that assumes two phases of gas accretion and star formation efficiency.  
%To interpret these observations, 
%We conduct a systematic search over a wide parameter space for a viable model that best matches the observed iron and Magnesium abundance distribution functions. 
%By systematic searching over a wide parameter space, 
We find that the data favor a three-phase SFH that consists of an initial starburst, followed by a rapid star formation quenching episode and a lengthy, quiescent secular evolution phase. 
The metal-poor, high-$\alpha$ bulge stars ([Fe/H]$<$0.0 and [Mg/Fe]$>$0.15) are formed rapidly ($<2\;$Gyr) during the early starburst. The density gap between the high- and low-$\alpha$ sequences is due to the quenching process. The metal-rich, low-$\alpha$ population ([Fe/H]$>$0.0 and [Mg/Fe]$<$0.15) then accumulates gradually 
through inefficient star formation during the secular phase. This is qualitatively consistent with the early SFH of the inner disk. Given this scenario, a notable fraction of young stars (age~$<5\;$Gyr) is expected to persist in the bulge. 
%The following quenching process 
%model suggests a two-phase star formation and enrichment history in which the predominant metal-poor, $\alpha$-enhanced ([Fe/H]$<-0.2$ and [Mg/Fe]$>0.2$) bulge stars are formed in a short timescale ($<2\;$Gyr) during an early starburst while the less numerous metal-rich, solar-$\alpha$ ([Fe/H]$>0$ and [Mg/Fe]$<0.15$) population accumulates slowly via low-state star formation after the starburst until today. These two distinct star formation phases are connected by a rapid early quenching process which is the key to reproduce the observed $\alpha$-dichotomy throughout the Galaxy. 
%We interpret the systematically higher [Mg/Fe] of bulge stars compared to the solar neighborhood as an indicator of an earlier onset of star formation quenching in the inner Galaxy that propagates outwards. 
Combined with extragalactic observations, these results suggest that a rapid star formation quenching process is responsible for bimodal distributions in both the MW’s stellar populations and in the general galaxy population and thus plays a critical role in galaxy evolution.
%Combining with extragalactic observations a heuristic link appears that a rapid star formation quenching process is responsible for both bimodality in stellar populations in our Galaxy and that in general galaxy population, implying a critical role it plays in galaxy formation and evolution.   

%Our results imply that an early quenching episode, which shapes the well-established bimodalities both in galaxies (active versus passive) and stars in our Galaxy ([$\alpha$/Fe]-enhanced versus solar-[$\alpha$/Fe]), is ubiquitous and critical in general galaxy formation and evolution.  
%Uncertainties of best-fit parameters that are calculated based on fitting quality of models are provided. 
\end{abstract}

%\keywords{}
\begin{keywords}
		The Galaxy: abundances -- The Galaxy: bulge -- The Galaxy: formation -- The Galaxy: evolution -- The Galaxy: stellar content -- The Galaxy: structure.
\end{keywords}

\section{Introduction}
\label{sec:intro}
The chemical compositions of stars are valuable fossil records that preserve --- in many cases --- the chemistry of the interstellar medium (ISM) from which the stars were born. Observations of large samples of stars formed at different epochs therefore allow one to unfold the chemical enrichment and star formation history (SFH) of the host system with unprecedented accuracy. The Milky Way (MW) is an ideal laboratory to conduct such analysis on a galaxy scale where chemistry observations of large numbers of individual stars are achievable.   
%One of the example is the Galactic bulge which can be uniquely studied in the Milky Way owing to our ability to resolve individual stars and is therefore of great importance 
These observations have relatively recently extended to the Galactic bulge (compared to, e.g., the solar neighborhood and halo), which is the only example of a massive galaxy center that can be studied in this way. Our bulge is therefore of great importance to understand the stellar population and formation history of galactic bulges in general. 

Early studies of MW bulge chemistry revealed the existence of multiple stellar populations with a wide range of metallicity, spanning $\sim-1$ to $\sim+0.5$ dex \citep{whitford1983,rich1988,mcwilliam1994}. 
Large effort has since been devoted to mapping the detailed metallicity distribution function (MDF) in the Galactic bulge using tracers such as Red Giant Branch stars \citep{zoccali2008,rich2012,uttenthaler2012,johnson2013,schultheis2017,rojas2017,garcia2018},
red clump stars \citep{hill2011,ness2013,zoccali2017,rojas2014}, and even subgiant and dwarf stars \citep{bensby2013}. 
%Although traced by stars at different evolution stages,
The MDF in the bulge as derived in different works shows noticeable difference in peak structure, described as either two strong [Fe/H] peaks \citep[with a supersolar component at $>+0.3$ and a subsolar one at $<-0.4$; e.g.,][]{hill2011,schultheis2017} or multiple, less strong peaks within the same metallicity range \citep[e.g.,][]{bensby2013,ness2013}.  
 
In addition to the MDF, $\alpha$-element abundances serve as important constraints on the SFH of stellar populations, in both our Galaxy and the general galaxy population. In particular, the relative abundance ratio of $\alpha$-elements to iron ([$\alpha$/Fe]) is a powerful diagnostic of star formation timescale, as rapid star formation can be recognized from an enhanced [$\alpha$/Fe] in stellar composition \citep{matteucci1994,thomas2010}. 
It has become increasingly clear that the bulge is composed of at least two distinct populations: a dominant one comprising old, metal-poor, and $\alpha$-enhanced stars, and a less numerous one comprising intermediate-age, metal-rich stars with solar-like [$\alpha$/Fe] \citep[e.g.,][]{babusiaux2010,hill2011,schultheis2017,rojas2019,queiroz2020}. %; Hasselquist et al. in prep.).   

A long-running debate persists over whether the $\alpha$-element enhancement in the old bulge stellar populations is identical with that of thick disk stars, such as those in the solar vicinity. 
The important physical question underlying this discussion is whether the bulge and thick disk share the same early SFH or not. Slightly different, perhaps systematically higher, [$\alpha$/Fe] abundances in the old bulge stars were suggested in some works \citep{cunha2006,zoccali2006,lecureur2007,fulbright2007,zasowski2019}. 
%A small fraction of these metal-poor, $\alpha$-rich stars in the bulge show chemical anomalies, suggesting an accretion origin \citep{recio2017,schiavon2017,fernandez2017}.
On the other hand, other studies found no significant difference \citep{melendez2008,alves2010}, evidence that favors a picture with no distinct chemical ``bulge'' structure besides the inner thick disk in the Galactic center \citep{fragkoudi2018,dimatteo2019}. 
In this sense, the Milky Way would be a pure disk galaxy with a buckled bar \citep[e.g.,][]{nataf2017}. 

On the theoretical side, impressive progress has been made 
to interpret these observed chemical distributions in the bulge.  
%observations of metallicity and $\alpha$ abundances of stellar populations in the bulge. 
\citet{matteucci1990} performed one of the pioneering chemical evolution modelling of the Galactic bulge, which traces not only the enrichment history of global metal content but also Fe, Si, Mg, and O individually.  
By comparing the prediction of various models with the observed MDF from \citet{rich1988}, the authors reached the conclusion that the bulge formed on a much shorter timescale ($<0.5\;$ Gyr) than the disk, with more efficient star formation and a flatter IMF. 
Following the observations of two distinct stellar populations in the bulge, many chemical evolution models have been developed to account for possible multiple formation channels or phases in the bulge \citep[e.g.,][]{bekki2011,grieco2012,tsujimoto2012,haywood2018,matteucci2019}.   
\citet{haywood2018} proposed a heuristic two-phase SFH connected by a rapid quenching episode to explain the density dip in the MDF and [$\alpha$/Fe] distribution functions ($\alpha$-DF) observed in the APOGEE survey \citep{majewski2017}. By exploring a broad radial range of the inner Galaxy ($R_{\rm GC} < 7\;$kpc), \citet{haywood2018} suggested that the bulge and inner disk are indistinguishable in terms of chemical properties and that their chemical evolution has followed a similar path.   

Although enormous progress has been made in modelling the chemical evolution in the bulge, most works have been conducted in a qualitative way, in which viable models are commonly %empirically
assessed based on visual inspection on the comparison with observations. %To our best knowledge, no study has 
%In this work, we aim to perform a systematic search for the best SFH model parameters %and determine the uncertainties of model parameters 
%based on a quantitative comparison between models and multi-dimensional chemical data. 
%assessment of fitting quality of each model to the data. 
In this work we aim to perform a quantitative analysis using recent chemical data from the APOGEE survey in order to reveal the detailed star formation and enrichment history of the Galactic bulge. 
The paper is organized as follows: We introduce the sample selection and the observed MDF and $\alpha$--DF in \S\ref{sec:data}. In \S\ref{sec:model} we describe the chemical evolution model used in this work and explain the systematic search for the best-fit model and the result of this search. %A detailed comparison between the model predictions and observations is also included in this section. 
More discussions about interpretation of the best-fit model and room for improvement of the model are presented in \S\ref{sec:discussion}. We briefly summarize our results in \S\ref{sec:summary}.

\section{Data}
\label{sec:data}

\subsection{Sample selection}
\label{sec:sample_selection}

Our sample is selected from the Apache Point Observatory Galactic Evolution Experiment survey \citep[APOGEE;][]{majewski2017}, an on-going core project of the Sloan Digital Sky Survey IV \citep[SDSS-IV;][]{blanton2017}. APOGEE provides high-resolution, near-infrared spectra for $\sim$5$\times$10$^5$ stars throughout the Milky Way’s bulge, disc, and halo \citep{zasowski2013, zasowski2017}, using custom spectrographs \citep{wilson2019} at 2.5~m Sloan Telescope at the Apache Point Observatory \citep{gunn2006}, and the 2.5~m Ir\'en\'ee du~Pont telescope \citep{bowen1973} at Las Campanas Observatory. Radial velocities, stellar parameters, and chemical abundances are determined from the spectra %\citep{holtzman2015}
, using custom pipelines described in \citet{nidever2015}, \citet{garcia2016} and \citet{jonsson2020}. In this work we use the parameters and abundances of a recent APOGEE internal release that includes reduced observations until November 2019 \citep[using the DR16 pipeline;][]{ahumada2020},
%SDSS date release (DR16, \citealt{ahumada2020}) 
and spectro-photometric distances based on the procedure described in \citet{rojas2017}. 

To remove stars with problematic spectra or unreliable parameter determinations, we adopt a signal-to-noise (SNR) cut of 60 and use the APOGEE spectroscopic flags\footnote{http://www.sdss.org/dr16/algorithms/bitmasks/} to select stars with EXTRATARG$==0$; the 1st, 4th, 9th, 16th and 17th bit of STARFLAG$==0$, {which correspond respectively to COMMISSIONING, LOW\_SNR, PERSIST\_HIGH, SUSPECT\_RV\_COMBINATION, and SUSPECT\_BROAD\_LINES}; and the 19th and 23th bit of ASCAPFLAG$==0$, {which correspond to METALS\_BAD and STAR\_BAD}. 
Since elemental abundance determinations in APOGEE tend to be less reliable at low effective temperature, {$T_{\rm eff}$}, we further exclude stars with $T_{\rm eff}<3200$~K. We have tested that our results do not change significantly when adopting an effective temperature cut at 3500~K for the sample selection.

In this work, we analyze the [Mg/Fe]-[Fe/H] abundance plane. We adopt magnesium ([Mg/Fe]) to trace the $\alpha$ abundance, as it is shown to be the $\alpha$ element most reliably measured by the pipeline, with least dependence on effective temperature, \citep[][J{\"o}nsson et al.~in prep]{jonsson2018,rojas2019}. 
{Note that the [Fe/H] and other elements to iron ratio, [X/Fe], of APOGEE stars are not populated when a \texttt{PARAM\_MISMATCH\_WARN} flag is set. This flag is activated when the difference between the [Fe/H] abundance determined from Fe spectral features and the [M/H] (the abundance of total metal content determined from the entire spectrum) exceeds 0.1$\;$dex. The reason for this discrepancy is not yet completely clear to the APOGEE team, but the stars for which this flag is set are generally metal rich ([Fe/H] $>$ 0.0) and cool ($T_{\rm eff}$ $<$ 4000 K). We obtain $15\%$ more stars when using [M/H] instead of [Fe/H] for the sample selection. Because stars with this mismatch are mostly metal rich, the sample selected in this work based on [Fe/H] is likely biased with underestimated surface density at the high metallicity end.   
%{Note that some APOGEE stars show considerable difference ($>0.1\;$dex) between the abundance of iron ([Fe/H]), measured from the Fe windows, and the abundance of total metal content [M/H], measured from global spectra fitting. %As a result, a PARAM_MISMATCH_WARN flag is set in the ELEMFLAG bitmask in the catalog. 
%In our bulge sample, $\sim15\%$ of stars show this mismatch in metallicity and are excluded in our analysis given that we are unsure which metallicity measurement is more robust in this case (and we do not have calibrated [Mg/Fe] measurements). Since these stars are mostly metal-rich, we caution the reader that the surface density of stars at high metallicity end in this work is probably underestimated. 
We will discuss the potential effect on our results in \S\ref{sec:model_bestfit}. }    

%These stars 
%The [Fe/H] of these stars are set to be invalid by the APOGEE pipeline, and therefore the [Mg/Fe] %determination is not available as well. Given this limitation, these stars are not included in our sample. Since these stars are mostly metal-rich, we caution the reader that the surface density of stars at high metallicity end in this work is probably underestimated. We will discuss the potential effect on our results in \S\ref{sec:model_bestfit}. } 
%\textcolor{teal}{[[GZ: This isn't quite true.  First, it's not just stars in the bulge. The [Fe/H] is not ``set to be invalid'' --- these stars do not have their calibrated abundance tags populated, but ASPCAP certainly produces values for them (and for [Mg/Fe]).  The other important datum is, how many stars are we talking about?  Need to quantify the effect.]]}

%This effective temperature criteria only excludes less than 2\% stars and does not significantly affect the chemical abundance distributions of the sample. 

We select stars in the Galactic bulge region on our side with Galactocentric distances within $3\;$kpc ($R_{\rm GC}<3\;$kpc). %Stars with estimated position on the other side of the Galaxy tend to have less reliable distance determinations and are excluded in our sample. 
Since the APOGEE sampling is irregular at different heights above the mid-plane, we focus on the mid-plane ($|Z|<0.5\;$kpc), where the observations tend to be distributed more homogeneously. Our stellar sample on the mid-plane contains 4454 stars. To obtain the integrated abundance distribution for the whole bulge region, we adopt the metallicity- and $\alpha$-dependent scale height derived by \citet[][see more details in \textsection2.3]{bovy2012b}.

\begin{figure*}
	\centering
	\includegraphics[width=0.95\textwidth]{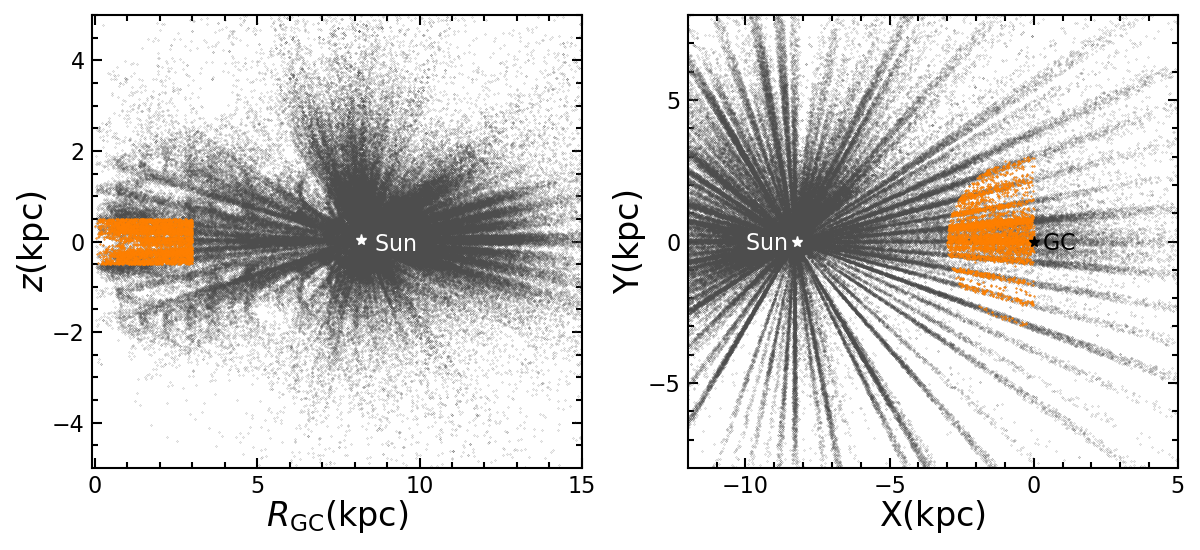}
	\caption{Spatial distribution of stars in the $R_{\rm GC}-Z$ plane (left-hand panel) and $X-Y$ plane (right-hand panel). APOGEE stars meeting the flag and SNR requirements as described in the text (\S\ref{sec:sample_selection}) are indicated by gray dots; stars in our final bulge sample with further spatial and effective temperature limits are denoted as orange dots.}
	\label{xy-rz}
\end{figure*}

Figure~\ref{xy-rz} shows the spatial distribution of APOGEE stars (grey dots) and our sample (orange dots) in the $R_{\rm GC}-Z$ plane (left-hand panel) and the $X-Y$ plane (right-hand panel). The positions of the Sun and the Galactic center are marked for reference.

%\subsection{[Fe/H]-[Mg/Fe] plane}
\subsection{Sample abundance distribution in the plane}
\label{sec:sample_abundances}

\begin{figure*}
	\centering
	\includegraphics[width=0.95\textwidth]{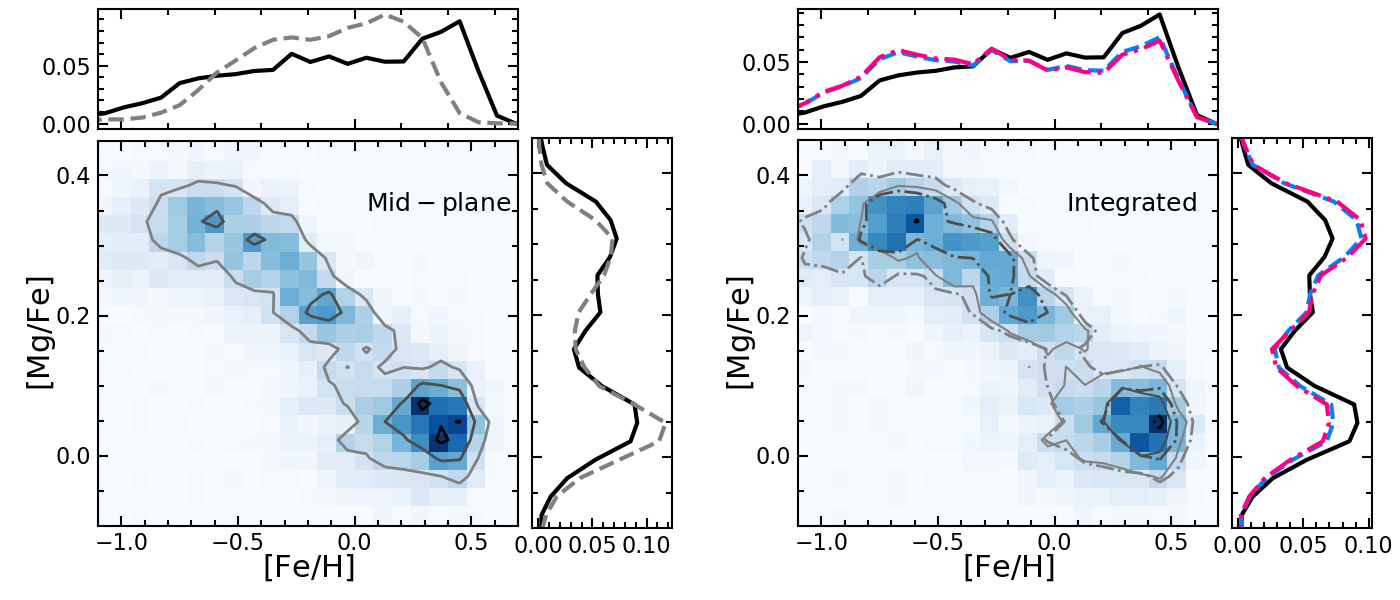}
	\caption{Density distribution of [Mg/Fe]--[Fe/H] for stars on the mid-plane (left-hand panel) and for the whole bulge region (right-hand panel). The projected distribution functions in [Fe/H] and [Mg/Fe] are normalized to the total number of stars and shown at the top and right sub-panels, respectively. Grey dashed lines in the left-hand sub-panels indicate the MDF and $\alpha$--DF in the inner disk taken from \citet{lian2020b}. %Lian et al. in prep. 
	In the right-hand panel, the dash-dotted and solid contours outline the integrated density distribution with and without consideration of disk flare, respectively. The red dash-dotted and blue dashed curves in sub-panels denote the corresponding MDF and $\alpha$-DF, respectively. Black solid curves are the original MDF and $\alpha$-DF on the mid-plane for reference.   
	%A clear bimodal distribution is present in both the distribution on the mid-plane and integrated one with a density gap at solar metallicity and [Mg/Fe]$\sim0.15$.}
	}
	\label{mgfe-feh}
\end{figure*}

The left panel of Figure~\ref{mgfe-feh} shows the observed density distribution of [Mg/Fe]--[Fe/H] for the sample on the mid-plane of the bulge region.  % for the mid-plane in the Galactic bulge region. 
A bimodal distribution is present, with a density gap at solar metallicity and $\rm [Mg/Fe] \sim 0.15$, suggesting at least two phases of star formation in the Galactic bulge. This gap in the [Mg/Fe]--[Fe/H] plane was also reported in \citet{rojas2019} based on APOGEE DR14 data. The metal-rich and metal-poor populations follow two [Mg/Fe]--[Fe/H] sequences with a small vertical offset in [Mg/Fe], consistent with the finding in \citet{hill2011}. The top and right sub-panels of Figure~\ref{mgfe-feh} show the projected distribution functions in [Fe/H] and [Mg/Fe] (the MDF and $\alpha$-DF), respectively, in black solid lines. The grey dashed lines indicate the MDF and $\alpha$-DF of the inner disk ($4<R_{\rm GC}<8\;$kpc) taken from \citet{lian2020b}.
%Lian et al. submitted. 
The bulge MDF shows a strong, wide peak at [Fe/H]$\sim$0.37$\;$dex, while the $\alpha$-DF exhibits a clear bimodal distribution, with two peaks at [Mg/Fe]$\sim0.32$ and $\sim0.05\;$dex, respectively and an interesting bump at [Mg/Fe]$\sim0.2$. 

A similar bimodal distribution has been found across a large portion of the disk from the inner Galaxy to the solar vicinity \citep{nidever2014,hayden2015}. %; Lian et al. in prep).%,Lian_2020_innerdisc}. 
The metal-poor, high-$\alpha$ branch is nearly invariant in chemical abundance space throughout the Galaxy except that the density distribution is more extended in both [Fe/H] and [Mg/Fe] and peaks at a slightly lower [Fe/H] in the bulge than in the solar vicinity. %($\rm \Delta[Mg/Fe] \sim 0.05$~dex).
In contrast, the metal-rich, low-$\alpha$ branch shifts systematically toward lower metallicity with increasing Galactic radius. This radial variation pattern suggests that the Galactic bulge and disk share a common early star formation history but follow different evolution paths in the more recent past, when the chemical thin disk (i.e., the low-$\alpha$ branch) formed.

\subsection{Integrated abundance distribution}

It has been shown that the vertical structure of stellar populations is dependent on their metallicity and [$\alpha$/Fe] \citep[e.g.,][]{bovy2012b,bovy2016}. More metal-poor and $\alpha$-enhanced stellar populations tend to have larger scale height and therefore dominate the stellar density distribution
%are located preferentially 
at larger vertical distance from the mid-plane.  
%\citet{bovy2012b} derived the spatial structure of ([Fe/H] and [$\alpha$/Fe]) mono-abundance sub-populations in the disk based on SDSS/SEGUE data. 
In the APOGEE survey, pointings targeting the inner Galaxy do not provide complete vertical or even angular coverage, but are localized to certain sky regions by design. 
The effect is that the observational sampling of the APOGEE survey at different heights in the bulge region is complex and irregular. 
To derive abundance distributions of the global bulge region, simply summing up APOGEE observations at different heights, given the significant vertical abundance gradient \citep{garcia-perez2018},
%\citep{hayden2014}, 
will introduce considerable bias for which it is difficult to account.  
%result in a biased abundance distribution that is not representative of the bulge region. 
%To avoid the potential bias introduced by the APOGEE target selection function, % in the abundance function, 

An effective alternative approach is to use the observed distribution in the mid-plane and account for the vertical distribution based on the known vertical structure of mono-abundance sub-populations. So far such mono-abundance analysis has only been conducted for the disk outside the bulge region \citep{bovy2012b,mackereth2017}.  
We therefore extrapolate the abundance-dependent disk scale height from \citet{bovy2012b}, taking the effect of the flare into account, to obtain the vertical structure of mono-abundance sub-populations in the bulge region. 
When no disk flare is considered, the scale heights of the high- and low-$\alpha$ populations ([Mg/Fe]$>0.15$ and [Mg/Fe]$<0.15$) are 562 and 233~pc on average, respectively.  
%an average scale height of 562 and 233$\;$pc are applied to the high- and low-$\alpha$ populations ([Mg/Fe]$>0.15$ and [Mg/Fe]$<0.15$), respectively.
For the flare, we adopt an average value of $R^{-1}_{\rm flare}=-0.1\;{\rm kpc^{-1}}$ for the low-$\alpha$ population \citep[see Fig.~12 in][]{bovy2016} and assume no flare for the high-$\alpha$ population as in that same paper. 
This choice of flare strength for the low-$\alpha$ population implies a 1.52 times lower scale height at $R_{\rm GC}=3\;$kpc compared to at the solar radius. 

For each [Fe/H]--[$\alpha$/Fe] bin, assuming an exponential disk in the vertical direction, the integrated surface density is derived following 
\begin{equation}
\label{eqn:integrated_density}
 d_{\rm tot}=\frac{d_{0.5}}{1-e^{-\frac{0.5}{h_{\rm Z}}}},
\end{equation}
where $d_{0.5}$ is the observed surface density within $|Z|<0.5\;$kpc and $h_{\rm Z}$ is the scale height of the given mono-abundance sub-population.

It is worth pointing out that this approach implicitly assumes a homogeneous vertical structure within the bulge region. It is known that the MW's bulge contains a buckled bar which complicates the situation. The Galactic bar barely affects the vertical distribution of metal-poor, high-$\alpha$ stars as evidenced by the consistent scale height ($\sim500\;$pc, \citealt{portail2017}) with their counterparts in the disk \citep{bovy2012b}. Therefore we assume that the structure of these stars near the solar neighborhood is applicable to the bulge. In contrast, the metal-rich, low-$\alpha$ stars in the bar are generally thicker than their counterparts in the disk due to bar buckling \citep{wegg2013}. Thus extrapolation of the structure of these stars from the disk could possibly underestimate their scale height in the bulge and therefore underestimate their integrated surface density. 

Note that the original abundance distribution on the mid-plane is equivalent to the integrated abundance distribution when assuming the same scale heights for all stars.  
Although the scale height of metal-rich, low-$\alpha$ stars in the bar is higher than that in the disk, it is still lower than the scale height of the metal-poor, high-$\alpha$ stars \citep{portail2017}. Hence taking the original observations on the mid-plane to represent the whole bulge would overestimate the scale height and thus the relative surface density of metal-rich, low-$\alpha$ stars. Therefore the original and integrated MDF and $\alpha$-DF could be considered as two extreme cases with overestimated and underestimated metal-rich, low-$\alpha$ stars, respectively.  
Since we lack information on the scale height of the mono-abundance population in the bulge, we use both the original and the integrated observations to constrain the bulge SFH as described in the next section. 
 
The right panel of Figure~\ref{mgfe-feh} shows the integrated density distribution of [Mg/Fe]--[Fe/H] and the projected MDF and $\alpha$-DF for the whole bulge region. The density map and solid contours show the integrated density distribution without consideration of the disk flare, while the dash-dotted contours outline the distribution with the disk flare taken into account. Comparing to the density distribution on the mid-plane (left panels), the integrated one exhibits a more prominent metal-poor, high-$\alpha$ branch. % density peak at [Fe/H]$\sim-0.6\;$dex.
This is because the denominator in Eq.~\ref{eqn:integrated_density} is larger for the metal-poor population, given their larger scale height compared to the metal-rich population. 

In Figure~\ref{mgfe-feh}, the integrated MDFs and $\alpha$-DFs with and without consideration of the disk flare are indicated as red dash-dotted and blue dashed curves, respectively. 
They are significantly different from their equivalent distributions in the mid-plane (black solid) with relatively more metal-poor, high-$\alpha$ stars due to those populations larger scale height. 
However, the integrated MDF and $\alpha$-DF barely change when taking the disk flare into account. 
%For both the mid-plane and whole bulge region, a clear [Mg/Fe]-bimodality is present which is much more stronger than that in MDF. 
There is an interesting bump in the $\alpha$-DF at [Mg/Fe]$\sim$+0.2, which extends the metal-poor, high-$\alpha$ branch in the MDF. The existence of this population reduces the significance of the bimodality in both the MDF and $\alpha$-DF. 

\section{Chemical evolution model}
\label{sec:model}

\subsection{Key ingredients}
\label{sec:model_ingredients}

To interpret observed chemical abundances in gaseous and stellar components of local galaxies, we developed a numerical chemical evolution model that accounts for three basic astrophysical processes involved in chemical enrichment: star formation, gas accretion, and galactic winds \citep{lian2018a,lian2018b,lian2018c,lian2019}. 
The model was further enhanced to be able to predict multiple elemental abundances of individual stars, in order to compare chemical evolution tracks with the enormous sets of stellar observations in the Milky Way (\citealt{lian2020a}, submitted). 
Here, we briefly introduce basic assumptions and inputs of the model and recommend
\citet{lian2018a} and \citet{lian2020a} to the reader for a more detailed description of the model development. 

We adopt the Kennicutt-Schmidt (KS) star formation law \citep{kennicutt1998} that connects the gas mass with star formation rate (SFR), but we allow the coefficient that regulates star formation efficiency (SFE: SFR/M$_{\rm gas}$) to vary (Table~\ref{tab:model_parameters}). This free coefficient is normalized to the original coefficient of the KS law (i.e., $2.5\times10^{-4}$). 
A \citet{kroupa2001} stellar initial mass function is adopted in the model. 
Considering the relatively high mass of our Galaxy, and the fact that metal outflow strength scales inversely with galaxy mass \citep{chisholm2018,lian2018a}, we assume no outflow in the model for the inner Milky Way. 

Metal production from asymptotic giant branch (AGB) stars, Type-Ia supernovae (SN-Ia), and core-collapse supernovae (CCSNe) are included in the model. 
Instead of assuming instantaneous mixing, we take the lifespan of stars fully into account, with 
metal-enriched material from evolved stars released to mix with ISM at the end of their evolution. 
We use metallicity-dependent AGB yields from \citet{ventura2013} and SN-Ia yields from \citet{iwamoto1999}. To study the effect of uncertainty in yields on our results, we use two sets of metallicity-dependent CCSNe yields from \citet[][hereafter CL04]{chieffi2004} and \citet[][hereafter K06]{kobayashi2006} and explore the best-fit model for each of them. 
To account for the time delay of SN-Ia since the birth of the SN-Ia-producing binary system, 
we adopt an empirical power-law SN-Ia delay-time distribution (DTD) with slope $-1.1$, which is calibrated to observed SN-Ia rates \citep{maoz2012}. %In \citet{lian2020a} we discussed the effect caused by various forms of SN-Ia DTD. 
We left the parameter $f_{\rm SN-Ia}$ free (Table~\ref{tab:model_parameters}), which regulates the global SN-Ia rate -- i.e., 
the fraction of stars within $3<M_*<16\;M_{\odot}$ that are SN-Ia progenitors -- to test whether it could be constrained by the abundance observations. {We adopt a minimum SN-Ia delay time of 35~Myr, which is widely used in chemical evolution models in the literature \citep[e.g.,][]{matteucci2009}. However, we will discuss the effect of adopting a different minimum SN-Ia delay time in \S\ref{sec:discussion_improvement} . } 
% mass loss due to stellar evolution have been taken into account in the model.  the  is assumed to take place at the life end of each star. 

%We note that the Magnesium abundance in the two CCSNe yields adopted here are underestimated relative to oxygen abundance. by $\sim$0.13 dex in K06 yields and $\sim0.15$ dex in CL04 yields (see more details in \textsection4). We increase the Magnesium production by the same amount in the corresponding yields. 

\subsection{Underestimate of magnesium production}
\label{sec:model_mg}

\begin{figure*}
	\centering
	\includegraphics[width=16cm]{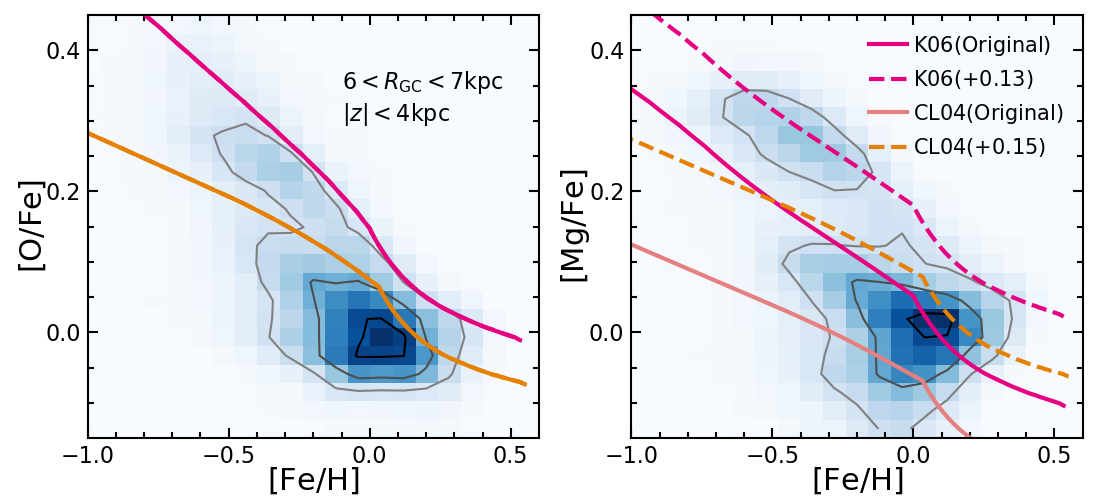}
	\caption{Prediction of models with original or Mg-enhanced CCSNe yields in [O/Fe]--[Fe/H] (left-hand panel) and [Mg/Fe]--[Fe/H] (right-hand panel). Solid lines indicate models with original CCSNe yields while dashed lines are models with Mg-enhanced yields. Observed stellar abundances in the disk within $6<R_{\rm GC}<7\;$ kpc and $|Z|<4\;$ kpc are shown in the blue density map and black contours.} 
	\label{mg-test}
\end{figure*}

When comparing observed $\alpha$ element-iron abundance ratios, [$\alpha$/Fe], with our chemical evolution models, we noticed evidence that the magnesium production in the K06 and CL04 CCSNe yields may be underestimated. 
To illustrate this, Figure~\ref{mg-test} shows the comparison between models with these yields and observed abundances in the [O/Fe]--[Fe/H] and [Mg/Fe]--[Fe/H] planes.
Observations within $6<R_{\rm GC}<7\;$kpc (also with APOGEE) are used in this comparison for two reasons. On one hand, the observed sample at smaller $R_{\rm GC}$ in the inner Galaxy is dominated by luminous stars for which APOGEE's oxygen abundances are in some cases less reliable \citep{jonsson2018,zasowski2019}. On the other hand, stellar populations in the outer Galaxy are intrinsically different from the bulge region, with far fewer high-$\alpha$ stars. As a compromise, stars within $6<R_{\rm GC}<7\;$kpc are used for the comparison. 
%Oxygen abundance tend to be less reliable for luminous stars which dominate the observed sample at inner Galaxy due to selection bias. However,   

%Comparing to the optical studies, magnesium is the most precisely and accurately measured element abundance in APOGEE \citep{jonson2020}. However, in the [Mg/Fe]-[Fe/H] plane, young dwarf stars seems to exhibit a peculiar `belly' feature at [Mg/Fe]=$-0.2$ and [Fe/H]=$-0.2$ (see Fig. 12 in \citealt{jonson2020}). This feature varies with different types of stars and is possibly due to  non-local thermodynamic equilibrium (NLTE) effect.    

We calculate a test model for both the K06 (magenta lines) and CL04 (orange lines) yields, with adopted parameter values listed in Table~\ref{para} as the ``Test model''. 
In Figure~\ref{mg-test}, it can be seen that at a given [Fe/H], the offsets between the models and data are clearly larger in [Mg/Fe] (right panel) than in [O/Fe] (left panel). For the model with K06 yields, when the model track in [O/Fe]--[Fe/H] is above the high-$\alpha$ branch, the track of the same model in [Mg/Fe]--[Fe/H] lies at lower [Mg/Fe]. 
We have confirmed that the offset in [Mg/Fe] with respect to [O/Fe] is not dependent on the choice of model parameters. 

{This systematic shift between different $\alpha$ elements is not likely to be due to systematics in the observed abundances. Based on the comparison with optical studies, oxygen and especially magnesium are both precisely and accurately measured in APOGEE
%; in particular, magnesium abundances has the highest precision and accuracy 
(see the discussion in \citealt{jonsson2020}). 
%Although [O/Fe] and [Mg/Fe] are well determined for the majority of stars, it is worth mentioning that a small number of stars seem to exhibit peculiar features in the [O/Fe]-[Fe/H] and [Mg/Fe]-[Fe/H] plane. For oxygen, systematic errors are likely present in giant stars with $T_{\rm eff}<4000\;$K and high [O/Fe] which give rise to a `finger' feature at high [O/Fe] and supersolar metallicity. For magnesium, young dwarf stars seems to exhibit a peculiar `belly' feature at [Mg/Fe]=$-0.2$ and [Fe/H]=$-0.2$ (see these features in Fig. 12 in \citealt{jonsson2020}). This `belly' feature varies with different types of stars and is possibly due to  non-local thermodynamic equilibrium (NLTE) effect. 
According to \citet{jonsson2020}, a small number of stars show signs of systematics in the abundance measurements; however, because these stars are concentrated within a limited range of chemical abundance, this behavior is unlikely to explain the discrepancy seen across a wide range of metallicity in Fig.~\ref{mg-test}. 
%This apparently can not explain the discrepancy at a wide range of metallicity as shown in Fig.~\ref{mg-test}. 
}

The most likely explanation for this discrepancy is that the magnesium production with respect to oxygen in both the K06 and CL04 CCSNe yields are underestimated. The underestimation of Mg in CCSNe yields has been highlighted in \citet{thomas1998}. A similar underestimated Mg/O ratio is also reported in other chemical evolution models in the literature \citep{sukhbold2016,rybizki2017,limongi2018}. 

{As a result, the chemical evolution model struggles to match the observed [Mg/Fe]--[Fe/H] distribution with the default CCSNe magnesium yields given the adopted minimum SN-Ia delay time of 35$\;$Myr. 
If no correction is applied to the Mg yields, then in order to reproduce the enhanced [Mg/Fe] of the high-$\alpha$ population, one needs to adopt a much higher minimum SN-Ia delay time. As we will discuss later on in \S\ref{sec:discussion_improvement}, with this modification and the original yields, the models cannot attain as good a match to the data.} 
%faces difficulties to match the relative fraction of low-$\alpha$, metal-rich stars. 
%push the model to the extreme, such as adopting a lower limit on the delay time of SN-Ia at $\sim$300~Myr. Note that this lower limit on SN-Ia delay time is much higher than that commonly adopted in chemical evolution models in the literature (e.g., 150 Myr in \citealt{andrews2017} and 35 Myr in Lian et al. in prep %\citealt{Lian_2020_innerdisc} 
%and this work) and is inconsistent with SN-Ia observations \citep{maoz2012}. 
%which marks the outset of rapid decrease of [Mg/Fe]. 
%Even more extreme conditions are needed to match models using the CL04 yields because of the lower intrinsic [Mg/Fe] in the yields table. Therefore, we consider that magnesium production in the CL04 and K06 yields is underestimated. 

In this work, we take the oxygen production as a benchmark to estimate the relative deficiency in Mg production. 
%We estimate the relative deficiency in Mg production in K06 and CL04 yields 
We calculate the net offset between the model tracks and the data in [Mg/Fe], with respect to [O/Fe], at a given [Fe/H]. The average offsets in K06 and CL04 yields are $-0.13$ and $-0.15\;$dex, respectively. We enhance the Mg production by the same offsets in these two CCSNe yields. The dashed lines in the right panel of Fig.~\ref{mg-test} show the model tracks with enhanced Mg production. With this modification, the models are able to match the observed oxygen and magnesium abundances simultaneously with physically plausible parameters. 
While we adopt a constant enhancement, the modification of magnesium production required to match the data is possibly metallicity dependent \citep{matteucci2020}. 
It is worth pointing out that the difference in magnesium production between the CL04 and K06 yields is much higher than the modification adopted in this work. Therefore we expect the effect of using different yields on our results would be much higher than the modification of magnesium production. 

%As described in \textsection3.1, the same enhancements are applied to derived the best-fit models in \textsection3.
%With CL04 yields, if no correction is applied to Mg yield we find no solution to reproduce the high-$\alpha$ branch in [Mg/Fe]-[Fe/H] based on our model. 

\subsection{Star formation framework}
\label{sec:model_framework}

To conduct a systematic search for the best-fit model, we adopt a simple and flexible star formation framework modified from the one in \citet{lian2020b}. %Lian et al. submitted %\citet{Lian_2020_innerdisc}
, in which we propose a novel multi-phase star formation framework to explain the joint distribution function of [Fe/H] and [Mg/Fe] in the inner disk ($4<R_{\rm GC}<8$~kpc). This framework consists of an early and a recent starburst that are associated with two relatively brief gas accretion events. 
%two short-lasting gas accretion phase associated with two starbursts are followed by another two prolonged phases of low-state star formation without gas supply. 
The early starburst is responsible for the formation of the chemical thick disk on a short timescale, while during the recent starburst, a metal-poor, moderately $\alpha$-enhanced population formed in the thin disk. %It is worth pointing out that this population is not observed in the Galactic bulge region which suggests that the Galactic center is not affected by the recent gas accretion and starburst event. 
Our model based on this multi-phase star formation history successfully reproduces the observed abundance distribution functions in the disk, especially the bimodal distribution in the [$\alpha$/Fe]--[Fe/H] plane. In this scenario, the well-known $\alpha$-bimodality arises due to a dramatically rapid decrease of star formation (i.e., star formation quenching) following the early starburst.  

%In light of \citet{Lian_2020_innerdisc}, 
We adopt a similar multi-phase star formation framework to interpret the abundance distribution functions observed in the Galactic bulge. 
Note that the metal-poor, moderately $\alpha$-enhanced population observed in the disk is absent in the bulge, suggesting that the inner Galaxy is not affected by the recent gas accretion event manifested in the disc. We therefore adopt a simpler version of the star formation framework presented in \citet{lian2020b}. %Lian et al. submitted
%\citet{Lian_2020_innerdisc} 
without the second gas infall and starburst event.  
%Given this star formation framework, 
There are four free parameters that characterize this star formation framework:
\begin{itemize}
	\item Initial coefficient of the KS law during the early starburst, $Coe_{\rm burst}$;
	\item Time when gas accretion is switched off, $t_{\rm acc,tran}$;
	\item Time when coefficient of the KS law changes, $t_{\rm SFE, tran}$;
	\item The coefficient of the KS law after the early starburst, $Coe_{\rm post-burst}$. 
\end{itemize}
The initial gas accretion rate is fixed to be 1.5$\;M_{\odot}{\rm yr^{-1}kpc^{-2}}$, and no gas accretion is assumed to occur after $t_{\rm acc,tran}$. 
%Two of them, the initial and post-starburst normalized coefficient of KS law (${\rm Coe_{\rm intial}, Coe_{\rm post-burst}}$), regulate the SFE during the initial starburst and suppression phase afterwards, respectively. Another two parameters determine the transition epoch of gas accretion, $t_{\rm acc,tran}$, and SFE, $t_{\rm SFE, tran}$. 
Thus in total, there are five free parameters in our model: four parameters describing the SFH plus $f_{\rm SN-Ia}$, the parameter regulating the global SN-Ia rate (\S\ref{sec:model_ingredients}). 

\subsection{Best-fit model} 
\label{sec:model_bestfit}

\subsubsection{Systematic search for the best-fit model}
Although abundance distribution functions have frequently been used to infer the bulge SFH \citep[e.g.,][]{haywood2018,matteucci2019}, the result is usually evaluated via a qualitative comparison to the data without a systematic, quantitative search across a range of possible solutions. % and estimate of uncertainties in the model parameters. 
One goal of this work is to determine the SFH of the Galactic bulge region in a quantitative way by exploring an extensive parameter space and assessing the robustness of the result. 

To this end, we create a 5-D parameter grid and run our chemical evolution model for each combination of the parameters. The range and step size of each parameter grid dimension are listed in Table~\ref{grid}. These values are chosen based on our experience to achieve a compromise between a wide coverage of parameter space and acceptable computation time. 
In total, 216,000 models are calculated for each of the CL04 and K06 CCSNe yields. 
Then, given the derived SFH and chemical evolution history for each model, 
we generate a mock stellar catalog of surviving stars, considering observational abundance uncertainties. 
A conservative uncertainty of 0.02~dex and 0.03~dex is adopted for [Fe/H] and [Mg/Fe], respectively, which is generally higher than the uncertainties in the data release catalog by a factor of $\sim$2. %in order to account for potential systematic errors.
Some of the parameters in the best-fit models change very slightly if different [Fe/H] and [Mg/Fe] uncertainties are adopted, but the main conclusions of this paper remain unchanged.   
%A non-constant age uncertainty that varies with stellar age as described in Lian et al. 2020a is used.  

The abundance distribution functions derived from these mock catalogs are then compared to the observed APOGEE catalogs to assess the goodness of match to the data. 
This assessment is performed with an unweighted quasi-$\chi^2$ diagnostic defined as: % , the sum of difference square in MDF and $\alpha$-DF as: 
\begin{equation}
\label{eqn:chi2}
\chi^2_v = \sum_{\rm i}\frac{(X_{\rm mod,i}-X_{\rm obs,i})^2}{N\bar{\sigma}^2},
\end{equation}
where the sum is calculated over all [Fe/H] and [Mg/Fe] bins ($i$), $X_{\rm mod}$ and $X_{\rm obs}$ are the values of the model and observed DFs in each bin, and $\bar{\sigma}$ is the average Poisson error over all these bins. The quantity $N$ stands for the degrees of freedom in the fitting, which equals the number of [Fe/H] and [Mg/Fe] bins (here, 50) minus the number of free model parameters (5), so for the results described below, $N=45$. 
We adopt a constant error $\bar{\sigma}$ in order to use the constraining power of both density peaks and troughs in MDF and $\alpha$-DF.  %where errors tend to be higher. 
%Note that the correction for vertical distribution, $d_{\rm tot}-d_{0.5}$, is much higher than the inherent observational errors estimated by Monte Carlo simulations and can be a major contribution of uncertainties to the global abundance distribution functions. 
%Since the uncertainties in the correction will be no more than the correction itself, we directly use the correction as the observational error to calculate the $\chi^2$ which is probably systematically underestimated.  
The model that minimizes this reduced quasi-$\chi^2$ is considered the best-fit model. Since a constant error is used in the $\chi^2$ calculation, it does not affect the choice of the best-fit model. 

\begin{table}
	\caption{Parameter ranges of the SFH model grid. Quantities in parentheses are the grid step size of each parameter.  The first four parameters are described in \S\ref{sec:model_framework} and the fifth in \S\ref{sec:model_ingredients}.
	}
	\label{grid}
	\centering
	\begin{tabular}{l c}
	\hline\hline 
	 Parameter & Range \\
	 \hline
     $Coe_{\rm burst}$ & 0.1-1 (0.1) \\
     $Coe_{\rm post-burst}$ & 0.01-0.28 (0.01-0.02) \\
     $t_{\rm acc,tran}$ &  0.1-1.5 (0.1) \\ 
     $t_{\rm SFE, tran}$ & 0.6-2.2 (0.2) \\
     $f_{\rm SN-Ia}$ & 0.016-0.030 (0.2) \\
     %Range &  & 0.01-0.28 (0.01-0.02) & 0.1-1.5 (0.1) & 0.6-2.2 (0.2) & 0.016-0.030 (0.2)  \\
     \hline
	\end{tabular}\\  
	\label{tab:model_parameters}
\end{table}

\subsubsection{The best-fit models}
\begin{figure*}
	\centering
	\includegraphics[width=16cm]{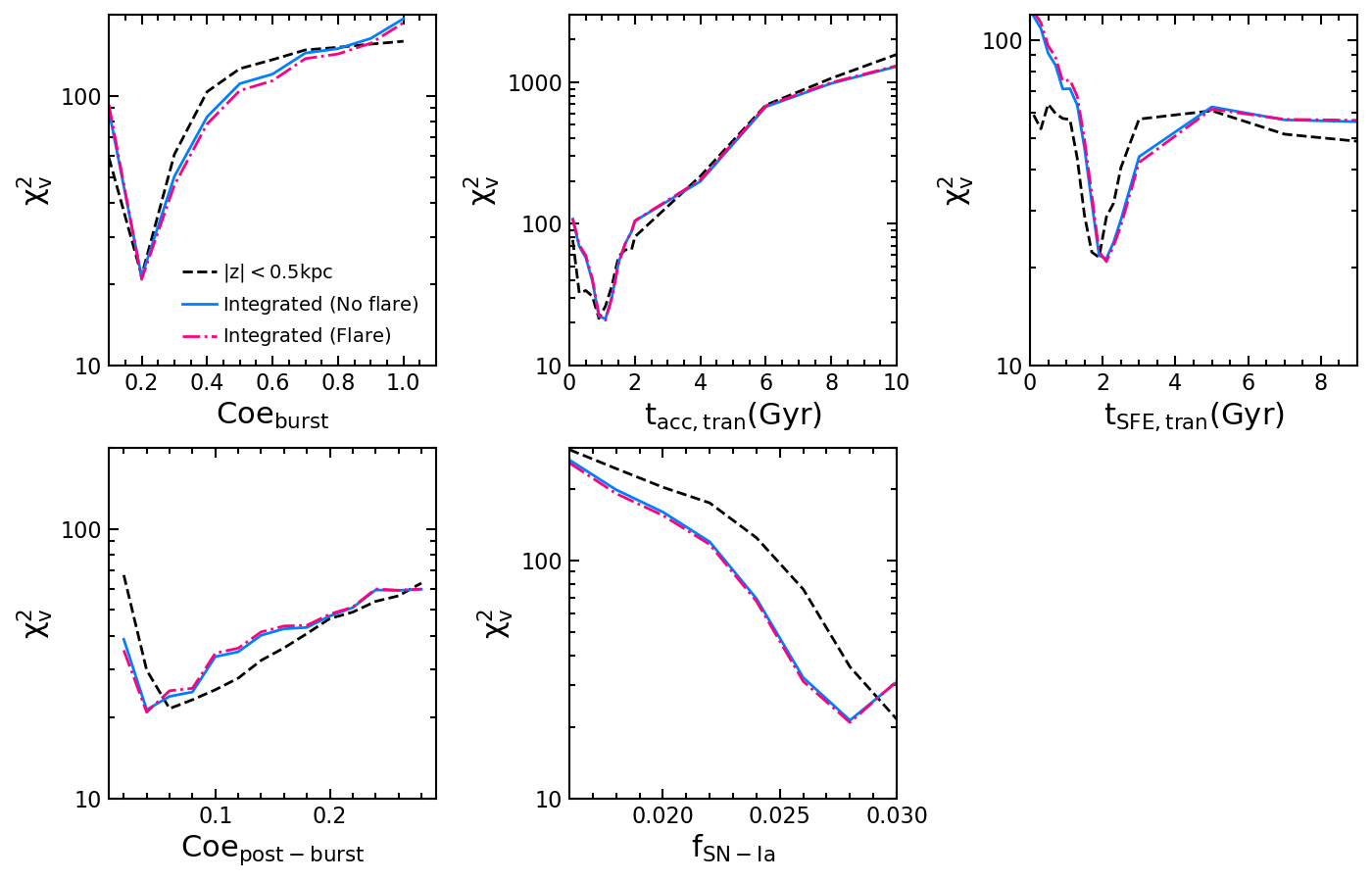}
	\caption{$\chi_\nu^2$ distribution as functions of each model parameter, with other parameters fixed to their best-fit values. The CCSNe yields from K06 are used in the models. Fitting results for observations in the mid-plane are shown as black dashed lines, and results for the integrated bulge region, with or without consideration of disk flare, are indicated as with magenta dash-dotted lines and blue solid lines, respectively. %The 32\% and 68\% probability of reduced $\chi^2$ value are marked as asterisks in each panel.}
	}
	\label{chi2-k06}
\end{figure*}

\begin{figure*}
	\centering
	\includegraphics[width=16cm]{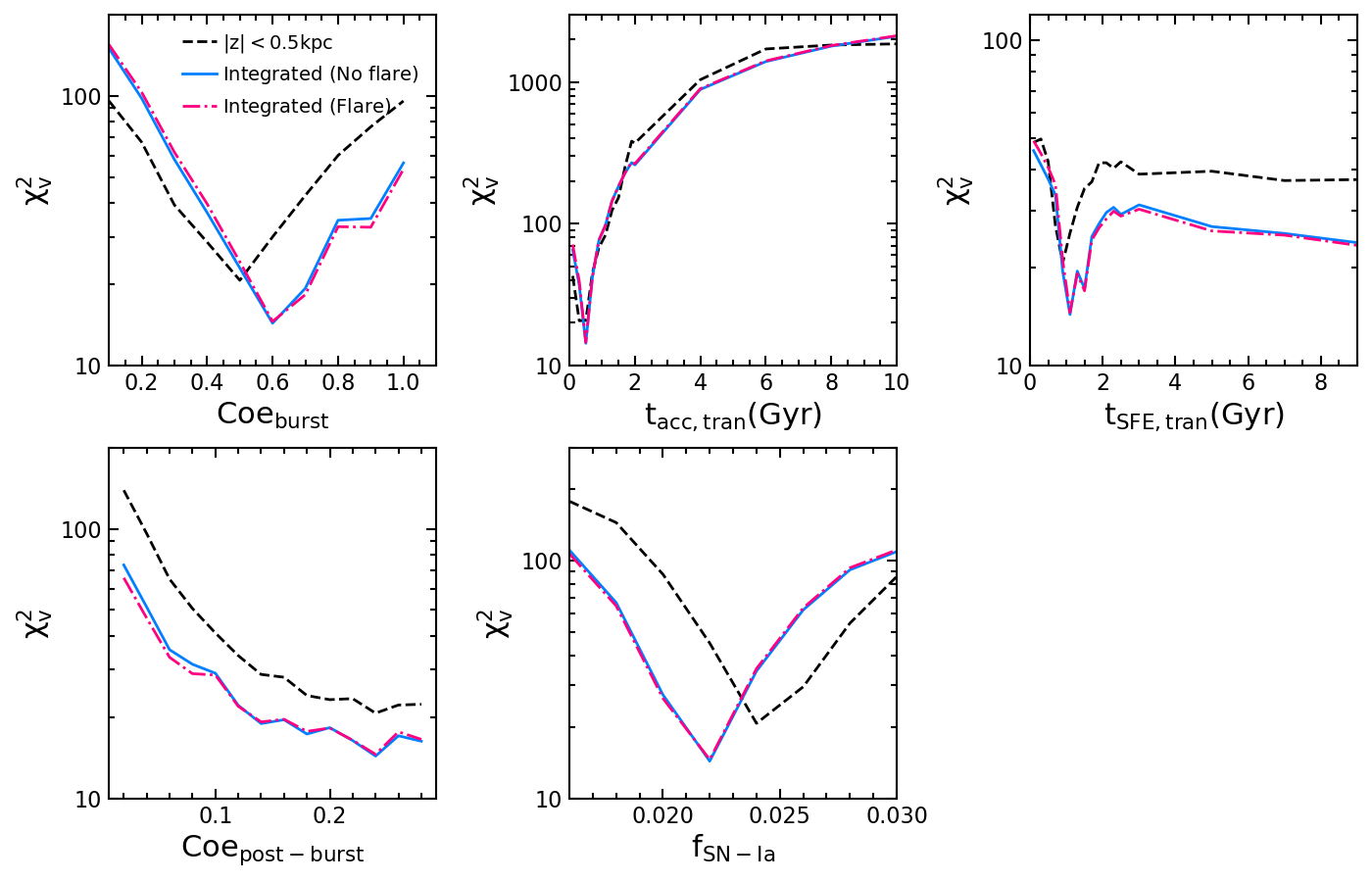}
	\caption{Same structure as Fig.~\ref{chi2-k06} but using CL04 CCSNe yields in the model.}
	\label{chi2-cl04}
\end{figure*}

Figure~\ref{chi2-k06} shows the $\chi_\nu^2$ as a function of each free parameter, with the other four parameters fixed to their best-fit values, for models with K06 yields. Figure~\ref{chi2-cl04} shows the equivalent results for the models with CL04 yields.
These best-fit parameters and minimized $\chi_\nu^2$ values are listed in Table~\ref{para}. 

%To illustrate how the best-fit parameters will change given different observations,
To illustrate the robustness of the fit, we include $\chi_\nu^2$ calculated for the abundance distribution functions on the mid-plane (black dashed line) and for the whole bulge region with and without considering the disk flare (magenta dash-dotted and blue solid lines, respectively). % and with correction considering disk flare (blue solid line). 
It can be seen that considering the disk flare barely affects the fitting results. This constancy is expected given the very small change in abundance distribution functions when taking the disk flare into consideration (see Fig.~\ref{mgfe-feh}). It also implies that the fitting results are robust against small variations in the data. 

A slightly different model, regardless of CCSNe yield choice, is preferred by the data on the mid-plane. 
As discussed in \textsection2, compared to the region high above the disk, the mid-plane in the bulge contains relatively more metal-rich, low-$\alpha$ stars and fewer metal-poor, high-$\alpha$ stars. 
%compared to global bulge, in the mid-plane there are 
As a result, models with less star formation in the initial starburst phase (i.e., lower $Coe_{\rm burst}$) or more star formation in the following prolonged secular phase (i.e., higher $Coe_{\rm post-burst}$) are favored by the unscaled mid-plane observations. 
However, the shape of the $\chi_\nu^2$ distribution of these fits are rather similar, which suggests the model parameters are consistently constrained by the observed abundance distribution functions without strong degeneracy. 
{Note that the sample selected in this work likely underestimates the surface density of the metal-richest stars ($\rm [Fe/H] > +0.2$; \S\ref{sec:sample_selection}), and thus the fraction of the metal-rich, low-$\alpha$ sequence. This implies more star formation in the secular evolution phase than the prediction of the best-fit models presented here. To account for that, a higher post-burst SFE would be required, and some residual gas accretion may also be needed to fuel more star formation. We expect the scale of change in these parameters would be comparable to the difference between the best-fitted models for the observations on the mid-plane and for the whole bulge region which also show considerably different fraction of metal-rich, low-$\alpha$ stars.}
%\textcolor{teal}{[[GZ: This last sentence is unclear.  To achieve what: a higher rate of late star formation?  Would that affect anything other than the surface density of metal-rich stars?]]}

Compared to K06 yields, the best-fit model based on CL04 yields has a relatively shorter initial accretion phase and therefore short starburst episode. Another difference is that the SFE is generally higher in the models with CL04 yields. These differences are mainly because [Mg/Fe] in the CL04 yields is $\sim0.3\;$dex lower than in the K06 yields. 
With this lower [Mg/Fe] in CCSNe yields, a higher star formation rate is needed to reproduce the observed [Mg/Fe] at a given [Fe/H]. More star formation results in more rapid enrichment and therefore reaching the density gap in [Fe/H] earlier, when the transition of star formation occurs. {Although the best-fit values of the parameters are different, the qualitative pictures of the two best-fit star formation scenarios are the same, and the fit quality to the data is comparable. More discussion on this is presented in \S\ref{sec:discussion}.} 

Given the consistent model parameters obtained for the original data on the mid-plane and the integrated ones for the whole bulge, hereafter we focus on the two best-fit models, one with K06 yield tables and one with CL04 yields, for the integrated abundance distribution in the bulge that accounts for the disk flare. 
%with vertical structural correction that includes the effect of disk flare in discussions later in this section. 

\subsubsection{Evolutionary history of the best-fit models}

%Following the approach described above, we obtain the best-fit models for each adopted CCSNe yields and observed abundance functions with correction for vertical distribution considering disk flare, with correction not considering disk flare, and without correction. The best-fit parameters are listed in Table~\ref{para}. 
%Given the K06 yields, the best-fit model does not change no matter whether considering disk flare or not. When focusing on the mid-plane without correction for vertical structure, a less dramatic drop in SFE after the initial starburst is needed. This is because of the lower number ratio between the high-$\alpha$ over low-$\alpha$ stars in the mid-plane. 
%the correction enhances the relative stellar density of high-$\alpha$ population over low-$\alpha$ population 
% are formed during the initial starburst and the stage afterwards, respectively. 

\begin{figure*}
	\centering
	\includegraphics[width=16cm]{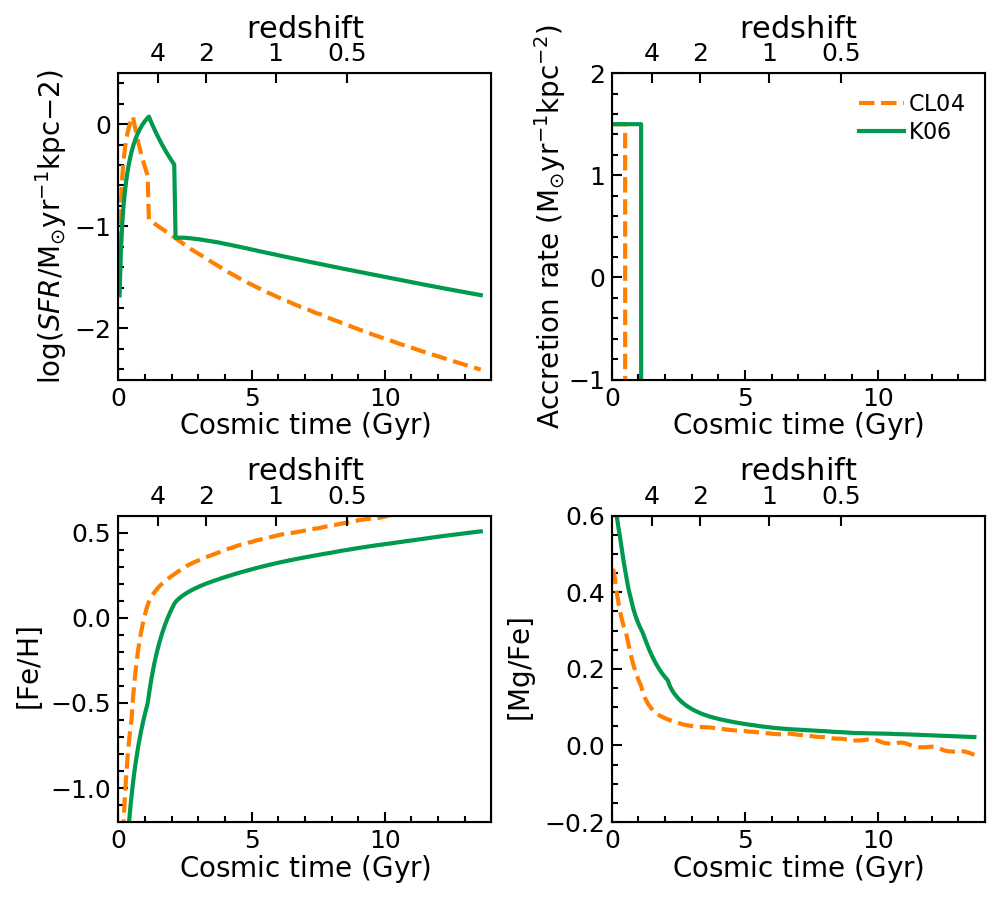}
	\caption{SFR (upper-left panel), Gas accretion (upper-right panel), [Fe/H] (bottom-left panel), and [Mg/Fe] (bottom-right panel)  histories of best-fit models based on CCSNe yields from K06 (solid green) and CL04 (dashed orange). 
	%Evolutionary histories of best-fit models based on CCSNe yields from (\citealt{kobayashi2006},  magenta) and (\citealt{chieffi2004}, green) in gas accretion (left-hand panel), SFR (left-middle panel), [Fe/H] (right-middle panel), and [Mg/Fe] (right-hand panel).}
    }
	\label{sfh}
\end{figure*}

Figure~\ref{sfh} shows the SFR (upper-left panel), gas accretion (upper-right panel), [Fe/H] (bottom-left panel), and [Mg/Fe] (bottom-right panel) histories of the two best-fit models based on CL04 (orange) and K06 (green) CCSNe yields. The initial gas accretion episode in both models lasts 1$\;$Gyr or less, suggesting the inner Galaxy has evolved largely as a closed-box system.   
There are two major phases of star formation: a short initial starburst associated with the initial gas accretion, followed by a prolonged secular evolution phase with low-level star formation activity. 
Similar to the disk model history in \citet{lian2020b}. %Lian et al. submitted,
%\citet{Lian_2020_innerdisc}, 
the early starburst is responsible for generating the metal-poor, high-$\alpha$ branch (i.e., the chemical/geometric inner thick disc), while the metal-rich, low-$\alpha$ branch is slowly built up during the later secular evolution. Therefore, the bulk of the old, metal-poor bulge stars is formed during the short-lived initial starburst.  
This rapid gas accretion and starburst in the early Universe that forms the bulge is consistent with the picture of coalescing giant clumps in a turbulent disk that has been proposed as a means of bulge formation in the literature \citep[e.g.,][]{elmegreen2008,inoue2012,clarke2019}.
Other channels can also trigger a short early starburst, including an initial dissipative collapse and/or early hierarchical mergers, which could possibly be involved in the bulge formation. 
%This starburst is possibly related to one or a combination of following astrophysical processes that have been proposed for bulge/thick disk formation on short timescale: , coalescence of gas clumps at high redshift \citep{zoccali2014,clarke2019}, or 

%Regardless of CCSNe yields used, 
An important episode shared by both modeled SFHs is the dramatic decrease in SFR right after the peak, which drops by one order of magnitude within 1~Gyr. This rapid SFR cessation (the so-called ``quenching'') process is critical to reproduce the observed  $\alpha$-bimodality \citep{haywood2018,lian2020b}. Interestingly, during this quenching episode, both models favor a two-step decrease of SFR with a relatively slower decline followed by a nearly straight drop. 
The first step is caused by the shutdown of the external gas supply (transition of gas accretion is earlier than the SFE).
%a decrease in the gas reservoir because of gas consumption by star formation and 
The second step, instead, is caused by switching the SFE to a lower values. 
This two-step cessation of SFR is required to explain the existence of the density bump in the $\alpha$-DF. % that stretches the MDF of the high-$\alpha$ population. 
The higher [Mg/Fe] of the metal-poor bulge stars, compared to stars in the thick disk in the solar vicinity, implies that this quenching process likely takes place earlier in the inner Galaxy and then propagates outwards, consistent with the inside-out quenching picture proposed to explain external galaxy observations \citep{ellison2018,wang2018,lin2019}. 
%We conclude that uncertainty in CCSNe yields mildly affects the adopted parameter values but does not weaken the constraining power of observed abundance distributions and alter the results of this work. 

%Although the picture given by the two best-fit models are essentially the same, there are some noticeable differences introduced by the different CCSNe yields. Compared to the model with Kobayashi06 yields, the model with CL04 yields tend to have slightly longer initial accretion phase and therefore longer starburst episode. Another slight difference between the SFH of these two models is the amplitude of the rapid SFR drop following the peak, which is $\sim0.15\;$ dex (or 30\%) lower in the model with CL04 yields. 

It is worth pointing out that the well-known bimodality of the general galaxy population in the color--magnitude (or color--mass) diagram---the blue cloud versus the red sequence---is also largely shaped by a rapid SFR cessation process that quiescent galaxies have experienced \citep[e.g.,][]{schawinski2014,lian2016}. This similarity to the quenching-induced $\alpha$-bimodality in the Milky Way suggests an intriguing link between the evolutionary path of our Galaxy and the general galaxy population, and that the SFR quenching process has been playing a critical role in shaping galaxies observed today, including our Galaxy. We offer a heuristic speculation that, if the recent gas accretion event that supplied metal-poor gas and boosted star formation in the disk (mostly in the outer disc; \citealt{lian2020a}) did not happen, our Galaxy would not appear to be star-forming today but instead be quiescent and red, like a S0 galaxy. 
%{An interesting direction worth deeper digging in the future.}
%a picture of two-phase star formation histories are 

\begin{table*}
	\caption{Best-fitting bulge SFH parameters (Table~\ref{tab:model_parameters}) for models based on CCSNe yields from \citet[][K06]{kobayashi2006} and \citet[][CL04]{chieffi2004}, fitted to observed abundances distributions on the mid-plane and in the integrated bulge region with and without considering the disk flare. 
	%with correction considering disk flare, with correction not considering disk flare, and without correction (no corr.). 
	The last column lists the reduced-$\chi^2$ value for each model. The parameters of the ``Test model'' discussed in \S\ref{sec:model_mg} and shown in Fig.\ref{mg-test} are also included. }
	\label{para}
	\centering
	\begin{tabular}{l c c c c c c}
		\hline\hline
		Model & $Coe_{\rm burst}$ & $Coe_{\rm post-burst}$ & $t_{\rm acc,tran}$ &  $t_{\rm SFE, tran}$ & $f_{\rm SN-Ia}$ & $\chi^2_v$ \\
		\hline
		& - & - & Gyr & Gyr & - & - \\
		\hline
	   %K06 (mid-plane) & 0.2$^{+0.01}_{-0.01}$ & 0.04$^{+0.05}_{-0.00}$ & 0.9$^{+0.20}_{-0.14}$ & 1.7$^{+0.14}_{-0.28}$ & 0.030 & 2.05 \\
	   %K06 (no flare) & 0.3$^{+0.03}_{-0.02}$ & 0.04$^{+0.03}_{-0.01}$ & 0.9$^{+0.13}_{-0.20}$ & 1.5$^{+0.18}_{-0.14}$ & 0.026$^{+0.0008}_{-0.0008}$ & 2.40 \\
	   %K06 (flare) & 0.3$^{+0.03}_{-0.02}$  & 0.04$^{+0.03}_{-0.01}$ & 0.9$^{+0.13}_{-0.20}$ & 1.5$^{-0.14}_{+0.23}$ & 0.026$^{+0.0008}_{-0.0008}$ &  2.47\\
       %CL04 (mid-plane) & 0.6$^{+0.09}_{-0.07}$ & 0.12$^{+0.08}_{-0.09}$ & 0.5$^{+0.07}_{-0.07}$ & 1.3$^{+0.27}_{-0.09}$ & 0.024$^{+0.0010}_{-0.0007}$ & 1.91 \\
	   %CL04 (no flare) & 0.6$^{+0.08}_{-0.08}$ & 0.12$^{+0.13}_{-0.05}$ & 0.5$^{+0.07}_{-0.07}$& 1.3$^{+0.22}_{-0.25}$ & 0.022$^{+0.0006}_{-0.0005}$ & 1.74 \\
	   %CL04 (flare) & 0.6$^{+0.09}_{-0.07}$ & 0.12$^{+0.10}_{-0.08}$ & 0.5$^{+0.07}_{-0.07}$ & 1.3$^{+0.27}_{-0.20}$ & 0.020$^{+0.0006}_{-0.0005}$ & 1.62 \\	
	   K06 (mid-plane) & 0.20 & 0.06 & 0.9 & 1.9 & 0.030 & 21.5\\
	   K06 (no flare) & 0.20 & 0.04 & 1.1 & 2.1 & 0.028 & 21.3 \\
	   K06 (flare) & 0.20 & 0.04 & 1.1 & 2.1 & 0.028 & 20.9 \\
	   CL04 (mid-plane) & 0.50 & 0.24 & 0.3 & 0.9 & 0.024 & 20.7 \\
	   CL04 (no flare) & 0.60 & 0.24 & 0.5 & 1.1 & 0.022 & 14.3 \\
	   CL04 (flare) & 0.60 & 0.24 & 0.5 & 1.1 & 0.022 & 14.6 \\  
	   \hdashline
	   Test model & 1.00 & 0.2 & 4.0 & 4.0 & 0.028 & - \\   	   
		\hline
	\end{tabular}\\  
%Notes $^a$: Difference as defined in Eq.~(2).\\
%$^b$: Test model shown in Fig.~\ref{mg-test} to illustrate the underestimate of Mg production in CCSNe yields. 
\end{table*}

\section{Discussion} 
\label{sec:discussion}

\subsection{A three-phase SFH for the bulge}
\label{sec:discussion_sfh}

\begin{figure*}
	\centering
	\includegraphics[width=16cm]{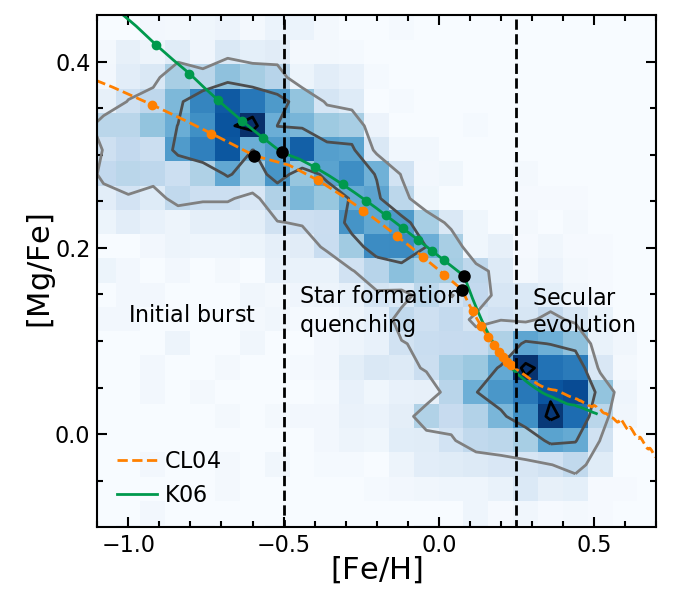}
	\caption{Integrated [Mg/Fe]--[Fe/H] distribution of the bulge region, overplotted with the evolution tracks of  best-fit models based on K06 (green solid) and CL04 (orange dashed) yields (\S\ref{sec:model_bestfit}). Orange and green circles on the model tracks mark  time intervals of 0.2$\;$Gyr for evolution in the first 2$\;$Gyr. Larger black circles highlight the important transitions of gas accretion and SFE in the model. 
	Vertical dashed lines separate the three primary phases of star formation, as discussed in the text.
	%Given the best-fit models, [Mg/Fe]--[Fe/H] diagram could be separated into three sections as indicated by vertical dashed lines: the metal-poor, high-$\alpha$ branch built in the early starburst, the in-between density gap due to a rapid star formation quenching process, and the metal-rich, low-$\alpha$ population gradually built through inefficient star formation during the long-term secular evolution stage. 
	}
	\label{track}
\end{figure*}

The SFHs of our best-fit models comprise three phases: an initial starburst, followed by a rapid star formation quenching episode, and a long-term secular evolution phase with a low SFR. Figure~\ref{track} shows the integrated [Mg/Fe]--[Fe/H] distribution of the whole bulge region and the predicted evolution tracks of our best-fit models. Green solid and orange dashed lines indicate the K06- and CL04-based models, respectively, as in Figure~\ref{sfh}. The large black circles highlight important transition times in the models, when the gas accretion switches off and the SFE ramps down. Other circles along the model tracks mark constant time intervals of 0.2$\;$Gyr for the evolution in the first 2$\;$Gyr.

Given this three-phase SFH, the [Mg/Fe]--[Fe/H] diagram can be separated into three general regimes (vertical dashed lines in Figure~\ref{track}). The metal-poor, high-$\alpha$ branch is mostly formed during the initial starburst. The following density valley between the high- and low-$\alpha$ branches is due to a rapid star formation quenching process. The metal-rich, low-$\alpha$ branch is gradually built through low SFR in the long-term secular evolution phase. This interpretation suggests that the density gap between the two main branches represents a transition between two modes of star formation in the bulge, from violent, bursty star formation at early times to long-term, low-state star formation more recently. This transition is qualitatively similar to the transition of disk formation from the chemical thick to the chemical thin disk \citep{haywood2018,lian2020b} %,Lian_2020_innerdisc} 
except for an earlier onset in the inner Galaxy. 

%Both the CL04 and K06 models do not show a clear plateau at the low metallicity end in Fig.~\ref{track}. % as seen in many observations \citep[e.g.,][]{francois2004}. 
%Instead, these two best-fit models flatten at [Fe/H]$\sim-$1.5 to $-2$, lower than the observed position at [Fe/H]$\sim$-1 \citep{deboer2014}. This is because the model in this work is constrained by the observed MDF and $\alpha$-DF, which are dominated by stars more metal-rich than the stars in the plateau. An improved model that is constrained by the distribution function that includes metal-poor stars ([Fe/H]$<$-1) could recover the plateau better. 

\subsection{Comparison with earlier bulge SFH measurements}
\label{sec:discussion_comparisons}

To account for the complex MDF in the bulge, a multi-phase SFH framework is generally adopted in the literature \citep{tsujimoto2012,grieco2012,haywood2018,matteucci2019}. 
The metal-poor bulge stars are generally believed to build up quickly at early times, while the metal-rich stars formed with a relatively longer time scale \citep{tsujimoto2012,grieco2012}. 
\citet{haywood2018} argued that a multi-phase SFH involving a rapid quenching stage adopted from \citet{snaith2015} could both explain well the global [Mg/Fe]--[Fe/H] relation and broadly reproduce the $\alpha$-bimodality in the inner Galaxy. By comparing the model-predicted MDF and $\alpha$-DF with observations, they concluded that the star formation quenching process is critical to reproduce the density dip between the high- and low-$\alpha$ branches. This empirical result is qualitatively consistent with our quantitative result. 

However, the SFH proposed in \citet{haywood2018} is quantitatively different from the SFH in our best-fit model. The early starburst of \citet{haywood2018} lasts $\sim1\;$Gyr longer, and the quenching episode starts much later, than in our bulge SFH ($\sim4\;$Gyr versus $\sim1\;$Gyr after initial star formation). The model in \citet{haywood2018} also faces challenges reproducing the observed MDF and $\alpha$-DF in detail, such as the relative number of high- and low-$\alpha$ stars. 

An alternative scenario to explain the MDF in the bulge is a bursty SFH, as proposed by \citet{matteucci2019}. In this scenario, the bulge SFH consists of multiple, short-lived starbursts that are separated by non-star forming gaps. This SFH is shown to be able to reproduce multiple peaks in the bulge MDF \citep{bensby2017}. A strong prediction of this scenario is a complex $\alpha$-DF, possibly with numerous peaks, that is inconsistent with the clear double-peaked bulge $\alpha$-DF as shown in this and other works.  

In addition to chemical evolution modelling, a large number of bulge SFHs have been derived using different methodologies \citep{nataf2015}. Most previous work relying on chemical compositions or photometric age estimates suggest that the bulge is a generally old component \citep{zoccali2003,grieco2012}. However, the inferred age distribution of bulge planetary nebulae peaks around $\sim3\;$Gyr, suggesting the presence of a considerable fraction of young populations in the bulge \citep{buell2013,gesicki2014},
a finding consistent with other photometric and spectoscopic studies \citep[e.g.,][]{bensby2013,bernard_2018_bulgeSFH}, and also present in barred galaxies in cosmological simulations of \citep{fragkoudi2020}. 
%This is qualitatively consistent with the 
A recent study of spectroscopic ages of bulge stars (Hasselquist et al., in prep) also confirms a non-negligible fraction of young stars (age $\sim$ 2--5~Gyr) that are preferentially found in the plane. 
%\citep{bernard_2018_bulgeSFH}

\begin{figure*}
	\centering
	\includegraphics[width=16cm]{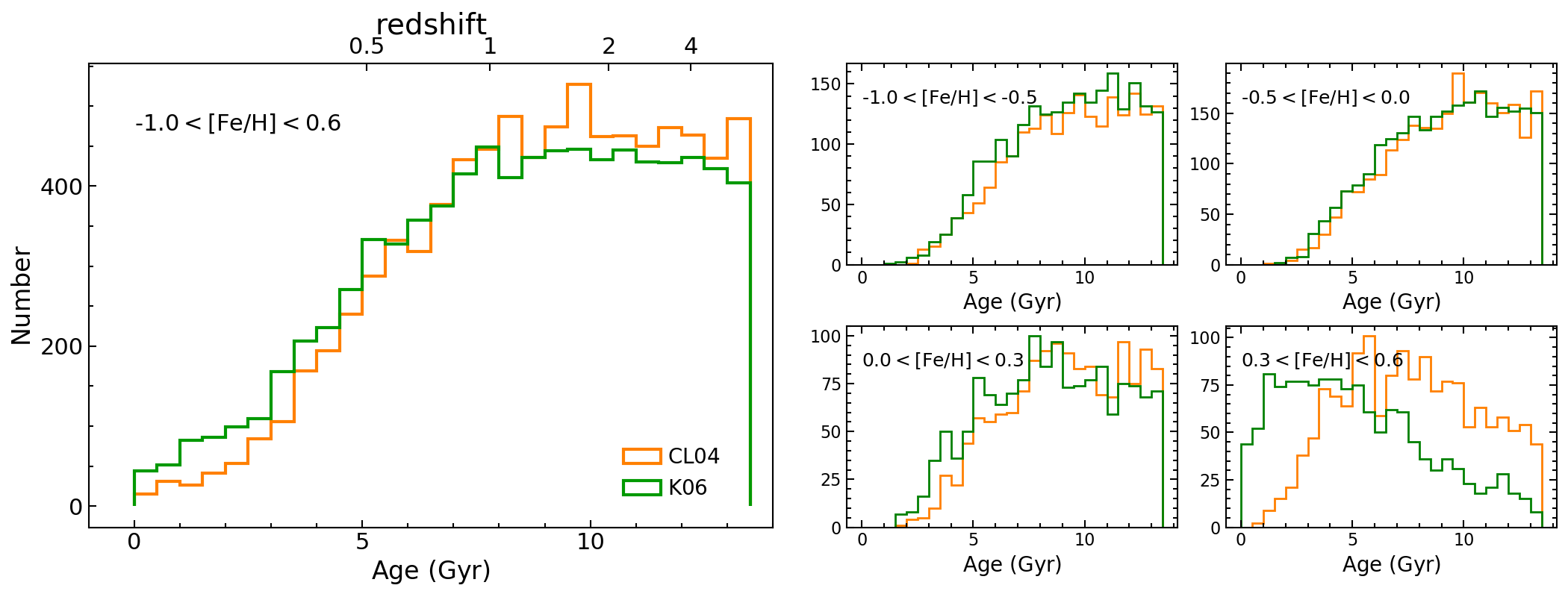}
	\caption{Global age distribution (left-hand panel) and that in four metallicity bins (right-hand panels) predicted by the best-fit models based on CL04 and K06 CCSNe yields. A notable fraction of young stars (age$<5\;$Gyr), which are preferentially very metal-rich, is present. 
	}
	\label{age-dtr}
\end{figure*}

The bulk bulge population predicted by our best-fit SFH is generally old, with a mass-weighted integrated stellar age of 10.5 and 9.7$\;$Gyr for models with CL04 and K06 yields, respectively. 
However, we note that a significant fraction of young stars is also predicted. Figure~\ref{age-dtr} shows the predicted age distribution of our best-fit models assuming an age uncertainty of 0.25$\;$dex. 
The left-hand panel indicates the global age distribution while the right-hand sub-panels show age distribution in four metallicity bins that correspond to the metallicity range of each episode in the best-fit SFH. The two intermediate metallicity bins correspond to the two stages of the star formation quenching phase.  %separated according to the three phases in the best-fit SFH as illustrated in Fig.~\ref{track}. 
The ages are derived based on a simple assumption that the star formation in the protogalaxy that would become the Milky Way started at 13.7$\;$Gyr ago. 
%there are a notable fraction of young stars with age$<5\;$Gyr, especially in the metal-rich population. 

Given an age uncertainty of 0.25$\;$dex, the number fraction of living young stars (age$<5\;$Gyr) predicted by the best-fit models are 11.3\% and 15.8\% for CL04 and K06 CCSNe yields, respectively. Note that this fraction is subject to change when different age uncertainty is assumed. For example, a considerably higher fraction of 25-29\% is expected when assuming an age uncertainty of 0.4$\;$dex. Comparing to the age distribution predicted by four distinct bulge SFHs as shown in Fig.~1 in \citet{nataf2015}, our result is more consistent with the age distribution derived by \citet{bensby2013}.

%Our result suggest that the fraction of young population inferred from chemical evolution modelling is dependent on the 

\subsection{Room for improvement in the model}
\label{sec:discussion_improvement}

\begin{figure*}
	\centering
	\includegraphics[width=18cm]{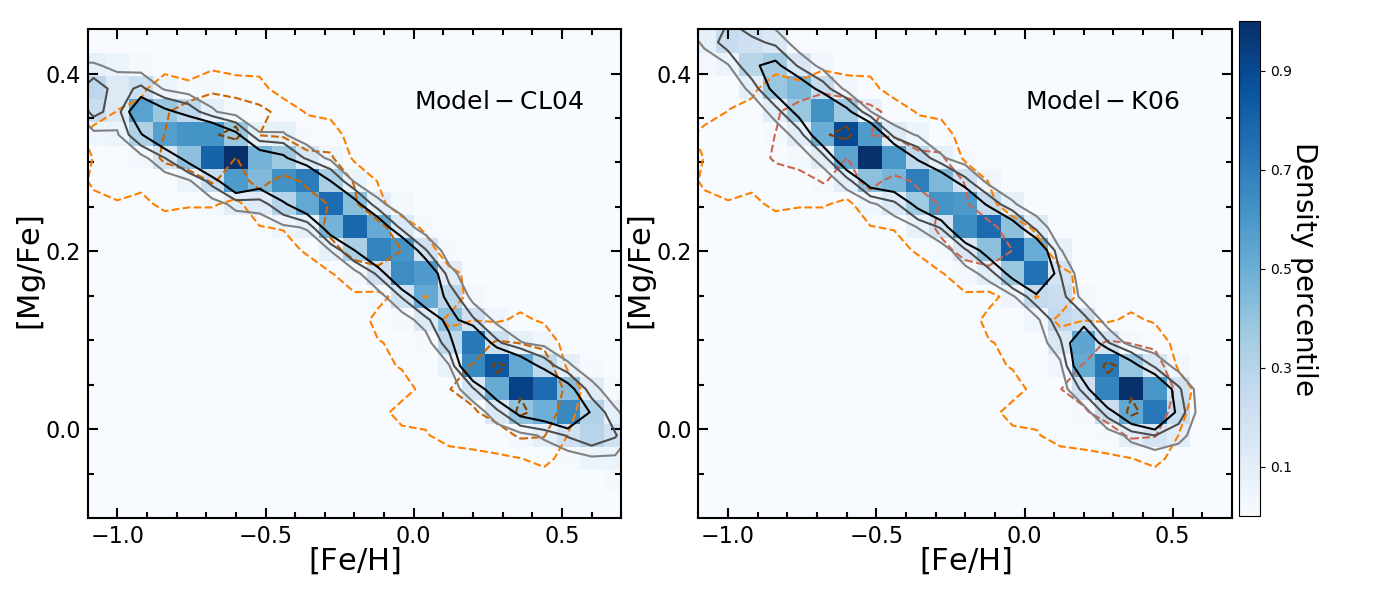}
	\caption{Density distribution of [Mg/Fe]--[Fe/H] as predicted by the best-fit models for CL04 yields (left-hand panel) and K06 yields (right-hand panel). The observed  distribution for the whole bulge region is overplotted as orange dashed contours for comparison.}
	\label{mgfe-feh-model}
\end{figure*}

\begin{figure*}
	\centering
	\includegraphics[width=16cm]{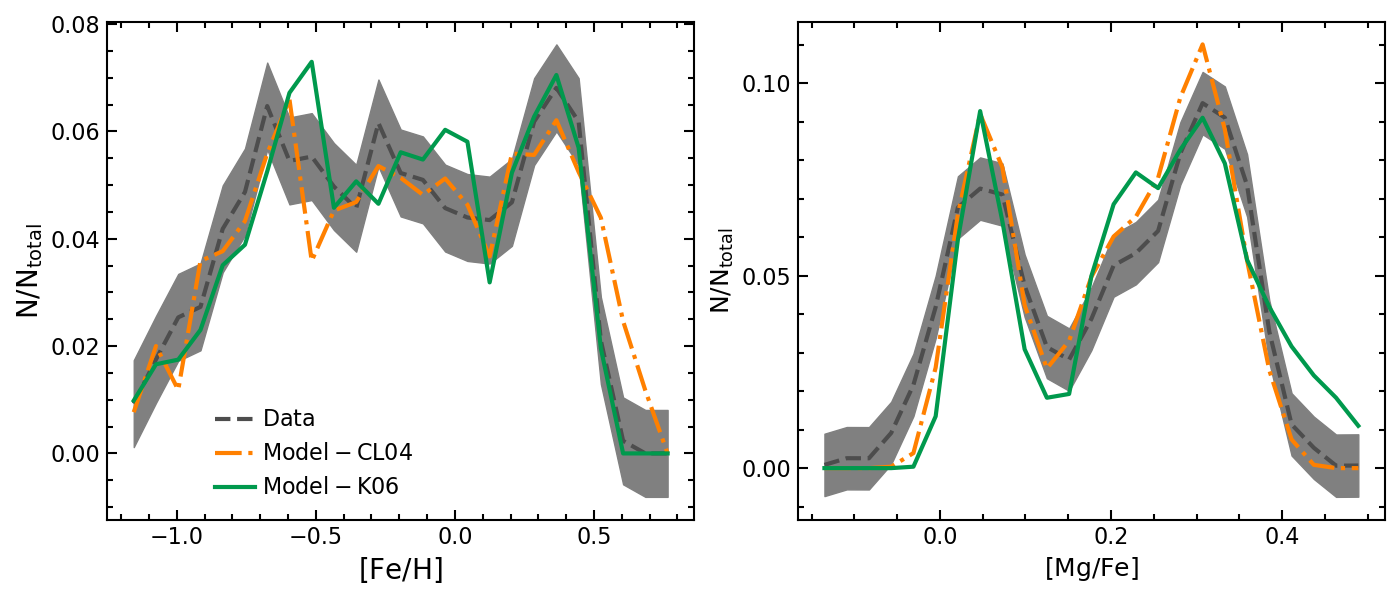}
	\caption{Comparison between the observed and modeled [Fe/H] (left-hand panel) and [Mg/Fe] (right-hand panel) distribution functions. The gray shaded region indicates the 3$\sigma$ scatter of the observed MDF and $\alpha$-DF, representing the average Poison uncertainty used to calculated the $\chi_\nu^2$ in Eq.~\ref{eqn:chi2}.
	%The error bar in the left panel indicates the average Poisson error of MDF and $\alpha$-DF which is used to calculated the $\chi_\nu^2$ in Eq. (2).}
    }
	\label{mdf-comp}
\end{figure*}

Figure~\ref{mgfe-feh-model} shows the density distribution of [Mg/Fe]--[Fe/H] as predicted by our best-fit models. The observed stellar distribution is shown as orange contours for comparison. 
Both models reproduce the global [Mg/Fe]--[Fe/H] trend and the $\alpha$-bimodality remarkably well. A noticeable difference is the narrowness of the tracks predicted by the models, compared to the data, in both low- and high-$\alpha$ branches. Figure~\ref{mdf-comp} compares the observed and predicted MDF and $\alpha$-DF for the best-fit models. The matches to the observed abundance distribution functions are equally good for these two models. The bimodal $\alpha$-DF is reproduced remarkably well, including the bump at [Mg/Fe]$\sim$0.2. The bimodality in [Fe/H] predicted by the models is also less significant than that in [Mg/Fe], with an extended tail at the metal-poor end, which is in good consistency with observations. 

Nevertheless, there remains room for improvement in the model fits to the abundance distribution functions. The low metallicity peaks in the MDF predicted by the models are slightly offset towards higher metallicity ($\sim-0.6$ versus observational $\sim-0.7$). This is a result of trade-offs to match the [Mg/Fe] distribution at the same time. 
Another under-performance in the models is that the predicted width of the low-$\alpha$ population in the [Mg/Fe] distribution is narrower than the data. This is also visible in Fig.~\ref{mgfe-feh-model}. One possible reason for this discrepancy could be that the observational uncertainties adopted in the model simulations are still underestimated. Another possibility is that the residual scatter is intrinsic, which suggests that the evolution path may not be smooth and unique but rather (at least somewhat) stochastic and inhomogeneous. The SFH implied by the best-fit models in this work could be considered as an average solution. To account for stochasticity in SFH requires a much more complex star formation framework, which is beyond the scope of this paper aiming at understanding the global bulge SFH. 

{Although our sample does not cover the low metallicity range ($\rm [Fe/H] < -1.2$), our best-fit models, especially the one with the K06 yields, keep increasing with decreasing [Fe/H] at $\rm [Fe/H] < -1$. This is not consistent with the observed flat [Mg/Fe] plateau at $\rm [Fe/H] < -1$ in many other previous works \citep[e.g.][]{mcwilliam1995}. 
The models also seem to not reproduce well the ``knee'' (change of slope) at $\rm [Fe/H]\sim-0.6$. 
%does not reproduce well the [Mg/Fe] plateau  and the . % the so-called `knee' feature in [$\alpha$/Fe]-[Fe/H] plane. 
%\textcolor{teal}{[[GZ: Possibly confusing---our sample doesn't cover this range but the models don't match the sample there?  Or are you asking readers to recall other studies of the plateau and knee?]]}
We hypothesize this mismatch is possibly due to the original [Mg/Fe] ratio in the CCSNe yields at low metallicity not matching the [Mg/Fe] value of the plateau, and/or the adopted minimum SN-Ia delay time being too low. 
}

\begin{figure}
	\centering
	\includegraphics[width=10cm,viewport=15 0 550 430,clip]{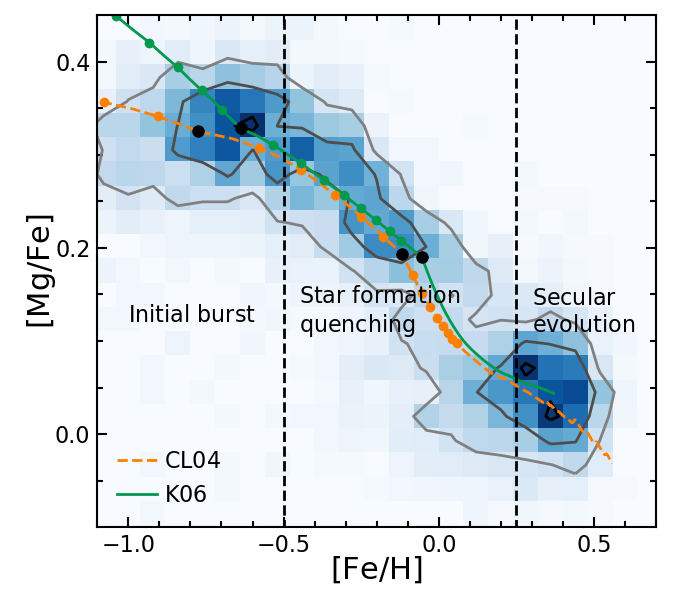}
	\caption{The same as Fig.\ref{track} except that a higher minimum SN-Ia delay time and default Mg production are assumed for the models.}
	\label{a1}
\end{figure}

\begin{figure*}
	\centering
	\includegraphics[width=16cm]{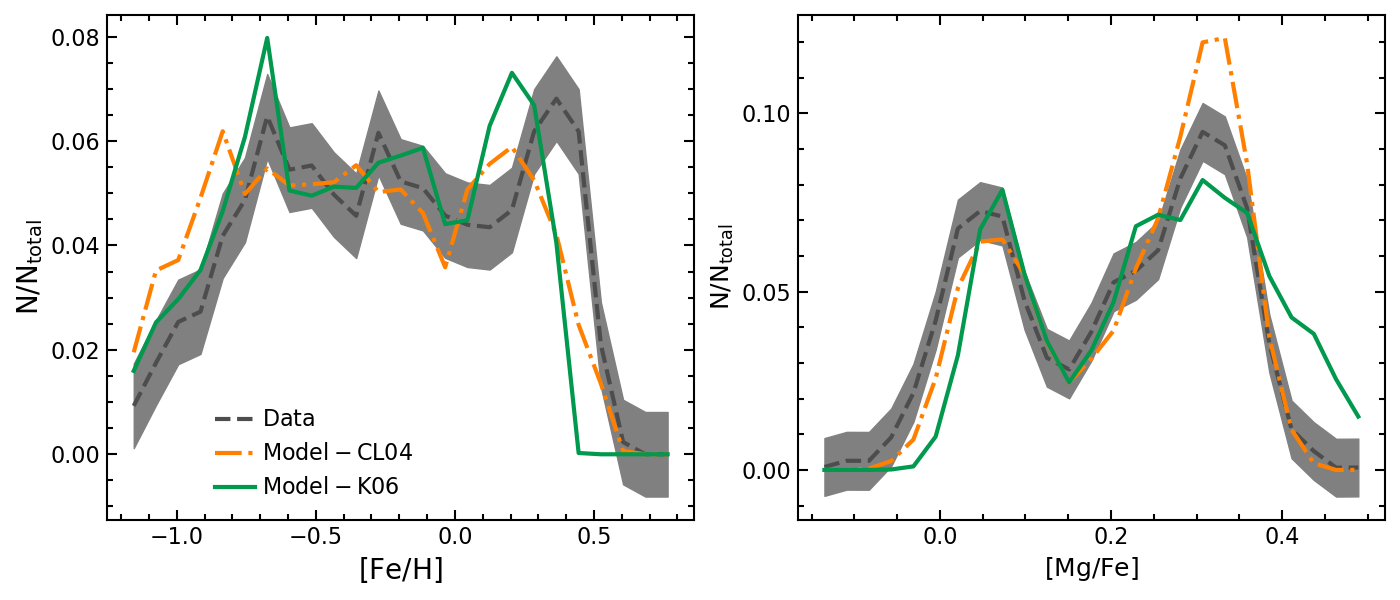}
	\caption{The same as Fig.\ref{mdf-comp} except that a higher minimum SN-Ia delay time and default Mg production are assumed for the models.}
	\label{a2}
\end{figure*}

{To test this latter idea, we calculated another set of test models, named `late-Ia' models here, for each of the CCSNe yields, adopting a higher minimum SN-Ia delay time (150$\;$Myr for CL04 and 70$\;$Myr for the K06 yields), the default yield tables (i.e., no Mg enhancement), and lower fractions of SN-Ia progenitors (0.012 for CL04 and 0.018 for K06 yields), which are required for the default Mg production to match the [Mg/Fe] of the low-$\alpha$ sequence. The value of minimum SN-Ia delay time is adopted for an example to demonstrate the impact of changing SN-Ia DTD, and all the other changes have to happen at the same time to maintain the [Mg/Fe]--[Fe/H] trend.  
%\textcolor{teal}{[[GZ: Tweaking all of these parameters to these values are necessary?  Is this just an example?  State clearly whether these are FITS (I'm guessing not) or examples to demonstrate the impact of changing the Ia DT and all the other changes that have to happen at the same time to maintain the density shape.]]}
The SFH of these late-Ia models are adopted as the same as the corresponding best-fit models.
%\textcolor{teal}{[[GZ: You already use ``test model'' to mean something else --- maybe ``late-Ia models''?]]}
Figure~\ref{a1} shows the evolutionary track of the late-Ia models in the [Mg/Fe]--[Fe/H] plane. The predicted MDF and $\alpha$-DF, compared with the observed data, are shown in Figure~\ref{a2}. It can be seen that these late-Ia models reproduce well the global [Mg/Fe]--[Fe/H] trend, and indeed show a more pronounced ``knee'' at $\rm [Fe/H]\sim-0.6$ and tend to flatten at lower metallicity.}

{However, the predicted abundance distribution functions do not match the observed ones as well as the best-fit models with a lower minimum SN-Ia delay time. The predicted MDF shifts systematically towards lower [Fe/H]. We have tested that the under-performance of late-Ia models remains when allowing SFH to vary.
%with higher minimum SN-Ia rate [[time?]]} is valid over the same model grid of the SFH. 
%\textcolor{teal}{[[GZ: What does it mean for an under-performance to be valid?  You mean, confirmed that these models consistently under-perform/have worse fits across the whole SFH grid?]]}
Since we focus on APOGEE observations in this work, we present the best-fit models to those data as our fiducial models. However, these late-Ia models illustrate one direction of future improvements when more observations of metal-poor stars are used to constrain the chemical evolution history.}    

\section{Summary}
\label{sec:summary}

We investigate the star formation history of the Galactic bulge by quantitatively modelling the [Fe/H] and [Mg/Fe] distribution functions simultaneously and matching to observations from the APOGEE survey. 
To avoid potential bias introduced by the varying sampling of the APOGEE survey at different vertical distances, we select our sample close to the mid-plane ($|z|<0.5\;$kpc) and account for the vertical structure of mono-abundance sub-populations as in \citet{bovy2012b,bovy2016} to infer integrated abundance distribution functions. A major effect of accounting for the vertical structure is an increase in the number of metal-poor, high-$\alpha$ stars with respect to the metal-rich, low-$\alpha$ stars because of the former's larger scale height. 

In both the raw and integrated [Mg/Fe] distribution functions, a clear bimodal distribution is present, with peaks at $+0.03$ and $+0.33\;$~dex and a gap at $+$0.15~dex. An interesting weak bump is also present at $+$0.2~dex. %Note that the high-$\alpha$ peak is shifted towards higher [Mg/Fe] by $\sim0.05\;$dex than in the solar vicinity. 
The integrated [Fe/H] distribution function exhibits three peaks at $\sim-0.67$, $\sim-0.27$ and $\sim+0.37\;$~dex with a dip at solar metallicity. 
%The distribution in [Fe/H] is also double-peaked at -0.70 and 0.38 dex with a gap at solar metallicity after considering the various vertical structure of mono-abundance sub-populations. 

To extract SFH information from abundance distribution functions, we apply our numerical chemical evolution model based on a star formation framework that allows single or multiple phases of star formation. 
To fully exploit the data, we explore a wide parameter space and systematically search for the best-fit model based on a quantitative assessment of fit quality. 
We obtain the best-fit model for the observations on the mid-plane as well as for the integrated ones across the global bulge region. 
%corrected for vertical distribution and also original data confined to the mid-plane. 
To test the effect of stellar yields on the results, we run two sets of models with different CCSNe yields (from K06 and CL04) and describe both sets of best-fit results.  

The best-fit models, regardless of the CCSNe yields used, 
suggest a three-phase SFH that consists of an early and intense starburst, followed by a rapid star formation quenching episode (one order of magnitude decrease in SFR within 1$\;$Gyr), and a prolonged stage of low star formation. The early starburst is responsible for the formation of the metal-poor, high-$\alpha$ branch; during the prolonged third phase, the metal-rich, low-$\alpha$ branch is gradually built up. These two star formation phases are connected by a rapid star formation quenching episode (an order of magnitude decrease in SFR within $\sim1\;$Gyr) that occurs in the early Universe. Because of the rapid decrease in SFR, the [Mg/Fe] ratio in the ISM also drops significantly and rapidly, leaving few stars formed with intermediate [Mg/Fe]. 
This quenching process is therefore critical to reproduce the $\alpha$-bimodality observed in the Milky Way, not only in the bulge region but also in the disk \citep{lian2020b}. 
%(Lian et al. submitted). %\citep{Lian_2020_innerdisc}. 
%The higher [Mg/Fe] in the bulge than the solar vicinity implies that the quenching process may have started earlier in the inner Galaxy and then propagated outwards. 
Future work will further address whether this quenching process took place simultaneously throughout the Galaxy or propagated inside-out or outside-in by studying radial variation of abundance distribution functions in more detail. 
Given the best-fit SFH in this work, a non-negligible fraction of young stars (age$<5\;$Gyr) is expected to exist in the bulge, a prediction that is broadly consistent with several recent age determinations of bulge stars \citep[e.g.,][Hasselquist et al., in prep]{bernard_2018_bulgeSFH}.  

There are mild differences between the best-fit models when using K06's or CL04's CCSNe yields. In general, the model with CL04 yields requires a higher SFE in both star formation phases to balance the lower intrinsic [Mg/Fe] in the yields table. However, the main characteristics of the three-phase SFH are not affected by the choice of CCSNe yields. We conclude that the uncertainty in CCSNe yields mildly affects the best-fit parameter values but does not significantly weaken the constraining power of observed abundance distributions on the bulge's SFH.  %and alter the results of this work. over
%We expect with progress in  stellar yields 

%\acknowledgements
\section*{Acknowledgements}
ARL acknowledge partial financial support to the APOGEE2-S survey through the QUIMAL project 130001 and FONDECYT project 1170476. DMN acknowledges support from NASA under award Number 80NSSC19K0589.

Funding for the Sloan Digital Sky Survey IV has been provided by the Alfred P. Sloan Foundation, the U.S. Department of Energy Office of Science, and the Participating Institutions. SDSS-IV acknowledges
support and resources from the Center for High-Performance Computing at
the University of Utah. The SDSS web site is www.sdss.org.

SDSS-IV is managed by the Astrophysical Research Consortium for the 
Participating Institutions of the SDSS Collaboration including the 
Brazilian Participation Group, the Carnegie Institution for Science, 
Carnegie Mellon University, the Chilean Participation Group, the French Participation Group, Harvard-Smithsonian Center for Astrophysics, 
Instituto de Astrof\'isica de Canarias, The Johns Hopkins University, Kavli Institute for the Physics and Mathematics of the Universe (IPMU) / 
University of Tokyo, the Korean Participation Group, Lawrence Berkeley National Laboratory, 
Leibniz Institut f\"ur Astrophysik Potsdam (AIP),  
Max-Planck-Institut f\"ur Astronomie (MPIA Heidelberg), 
Max-Planck-Institut f\"ur Astrophysik (MPA Garching), 
Max-Planck-Institut f\"ur Extraterrestrische Physik (MPE), 
National Astronomical Observatories of China, New Mexico State University, 
New York University, University of Notre Dame, 
Observat\'ario Nacional / MCTI, The Ohio State University, 
Pennsylvania State University, Shanghai Astronomical Observatory, 
United Kingdom Participation Group,
Universidad Nacional Aut\'onoma de M\'exico, University of Arizona, 
University of Colorado Boulder, University of Oxford, University of Portsmouth, 
University of Utah, University of Virginia, University of Washington, University of Wisconsin, 
Vanderbilt University, and Yale University.

\section*{Data Availability Statements}
The data underlying this article is from an internal incremental release of SDSS-IV/APOGEE survey, following the SDSS-IV Data Release 16. This incremental release is not publicly available now but will be included in the final public data release of SDSS-IV in 2021.   
%which are available at \url{https://data.sdss.org/sas/dr16/apogee/spectro/aspcap/r12/l33/allStar-r12-l33.fits}, with datamodel at \url{ https://data.sdss.org/datamodel/files/APOGEE_ASPCAP/APRED_VERS/ASPCAP_VERS/allStar.html}. 
The chemical evolution model results are available upon request. 
%\clearpage

%\bibliographystyle{apj}
\bibliographystyle{mnras}
\bibliography{jianhui}{}

%\appendix
%%\section{Test models for different minimum SN-Ia delay times}
%Here we present the predicted [Mg/Fe]--[Fe/H] evolution track and abundance distribution functions by the \textcolor{teal}{test model for which we adopt higher minimum SN-Ia delay time} than the best-fit models presented in \S\ref{sec:model_bestfit}. Figure~\ref{a1} shows the evolution track while Figure~\ref{a2} shows the predicted MDF and $\alpha$-DF. 
%\textcolor{teal}{[[GZ: Seems a bit odd to have the figures here in an appendix, while there are several paragraphs of text in the main paper.  (Especially since those paragraphs contain the model details needed to understand these figures.)  Suggest to either put the whole discussion here in the appendix, or move these plots to the main paper, depending on how much you want to emphasize this section (ultimately, how much you think the referee will object to a 35~Myr delay time?).]]}

\bsp	% typesetting comment
\label{lastpage}
\end{document}